\documentclass[preprints,review,accept,pdftex,moreauthors]{Definitions/mdpi} 
\firstpage{1} 
\pubvolume{1}
\issuenum{1}
\articlenumber{0}
\pubyear{2026}
\copyrightyear{2026}
\datereceived{ } 
\daterevised{ } 
\dateaccepted{ } 
\datepublished{ } 

\usepackage[utf8]{inputenc}
\usepackage{graphicx}
\usepackage{amsmath}
\usepackage{amssymb}

\usepackage{aas_macros}       

\usepackage{xcolor}

\definecolor{orange}{rgb}{1,0.5,0}
\definecolor{darkgreen}{rgb}{0,0.5,0}

\newcommand{\supcite}[1]{}

\newcommand{\HI}{\mbox{H{\scriptsize I}}\,} 
\newcommand{\mHI}{\mathrm{H_I}} 
\newcommand{\Tsys}{$T_\mathrm{sys}$}      
\newcommand{\mTsys}{T_\mathrm{sys}}      
\newcommand{\mMsol}{\mathrm{M_\odot} } 

\newcommand{\dang}{d_A}   

\newcommand{\LCDM}{$\Lambda \mathrm{CDM}$ }    

\Title{Cosmology with post-EoR 21 cm Intensity Mapping surveys: power spectrum measurements in the last decade} 

\Author{Réza  Ansari $^{1}$\orcidA{}, Xuelei Chen$^{2}\orcidA{}$, Albert Stebbins $^{3}$ and Peter Timbie $^{4}$\orcidD{}}

\AuthorNames{Réza  Ansari , Firstname Lastname and Firstname Lastname}

\address{%
$^{1}$ \quad Université Paris-Saclay,  
Université Paris Cité, CEA, CNRS, AIM,  
91191 Gif sur Yvette, France; reza.ansari@universite-paris-saclay.fr \\
$^{2}$ \quad State Key Laboratory of Radio Astronomy and Technology, National Astronomical Observatory, Chinese Academy of Sciences, 
20A Datun Road,  Beijing 100101, P. R. China; xuelei@bao.ac.cn \\
$^{3}$ \quad Fermi National Accelerator Laboratory, P.O. Box 500, 
Batavia IL 60510, USA; stebbins@fnal.gov \\
$^{4}$ \quad Department of Physics, University of Wisconsin - Madison, 
1150 University Ave, Madison WI 53703, USA ; pttimbie@wisc.edu 
}

\corres{Correspondence: reza.ansari@universite-paris-saclay.fr ; } 

\abstract{The technique of 21\,cm intensity mapping (IM) is emerging as a possible, cost-effective method to supplement 
optical surveys to map the large-scale structure, and extend their redshift reach, with projects aiming to 
cover a large range of post-EoR redshifts, from z=0 to z=6. 
Since the early discussions 20 years ago, a few dedicated interferometers have been built, and surveys have been carried out using
them as well as general purpose radio instruments. An introduction to late-time cosmology with 21\,cm intensity mapping is given, highlighting the main scientific and technical challenges. We present then an overview of different instrument concepts and 
survey strategies, followed by a discussion of outstanding data analysis challenges, most notably the foreground  subtraction, which is a major issue in IM. We carry on with a presentation of the currently active projects and the 21\,cm signal measurements which have been obtained, either in cross-correlation with optical surveys, 
or as the autocorrelation power spectrum. Finally, we discuss briefly the prospects for this field with future instruments and surveys such as SKA, HIRAX and CHORD.  
}


\begin{document}






\section{Introduction} 
\label{sec:introduction}
The standard model of cosmology, $\Lambda$CDM, posits a universe dominated by a cosmological constant ($\Lambda$) and cold dark matter (CDM) and a nearly scale-invariant spectrum of inhomogeneities generated in the early universe. This model provides a remarkably accurate description of a wide range of observations, particularly of the distribution of galaxies and gas as measured by optical galaxy surveys and the cosmic microwave background (CMB).  Small deviations from this idealized model can provide clues as to the nature of dark energy (which may be a cosmological constant) that drives the accelerated expansion of the universe, dark matter, which binds galaxies together, and the origin of the universe itself. There is very little known of these two substances nor of the very early history of our universe.  These cosmological mysteries are some of the most profound open questions in modern physics \citep{2012PhR...513....1C, 2013PhR...530...87W}.

A primary avenue for discovering subtle features of the universe not encoded in the idealized $\Lambda$CDM model is improved measurement of the large-scale structure (LSS) of the universe i.e.~of the distribution of galaxies and gas \citep{2011PhRvD..84f3505B, 2014JCAP...01..042D}. The 21\,cm line emission from neutral atomic hydrogen (\HI) provides a particularly good probe of LSS since this line is prominent in the radio band, which is unaffected by dust and reddening and where there is essentially no other line emission.
Thus, to a good approximation the wavelength of a spectral feature can be converted to a Doppler redshift without having to first identify the atomic transition.  The direct determination of redshifts using 21\,cm data can be compared to the corresponding optical technique, which requires identifying a suitable subset of target galaxies (photometry), then obtaining an optical spectrum, and finally finding some unique combination of emission and absorption lines that allows an unambiguous determination of the redshift for that galaxy (spectroscopy). Furthermore, \HI has existed in abundance throughout the history of the observable universe, back to the surface of last scattering at $z\sim10^3$, lasted through the dark ages ($20\lesssim z\lesssim10^3$), before being ionized by UV and X-ray photons once star formation started during the epoch of reionization (EoR). While optical techniques can only observe visible matter, 21\,cm emission and absorption occur even before galaxies form, i.e. during the cosmic dark ages.  

This paper is focused on measurements of \HI after the EoR when most \HI exists inside but not outside galaxies. There is a long history of 21\,cm galaxy redshift surveys
\cite{
ZwaanEtAl1997_AHISS,
SpitzakSchneider1998,
1999PASA...16...48R,
RosenbergSchneider2000_ADBS,
LangEtAl2003_HIJASS,
MinchinEtAl2003_HIDEEP,
MeyerEtAl2004_HICAT,
KoribalskiEtAl2004_BGC,
DonleyEtAl2005_HIZOA,
WongEtAl2006_NHICAT,
2018ApJ...861...49H,
ZhangEtAl2024_FASHI}
but here we consider a newer technique for mapping the cosmic distribution of neutral hydrogen gas (\HI), namely 21\,cm Intensity Mapping (IM) \citep{2008PhRvL.100i1303C,Loeb2008,Ansari2008}. 
IM refers to the detection of underlying structures formed by \HI-containing sources, such as galaxies, by mapping the intensity of \HI emission without requiring  detection of individual sources \citep{2001JApA...22...21B,2004MNRAS.355.1339B}, \citep{2008PhRvL.100i1303C, 2010ApJ...721..164S, 2012A&A...540A.129A}. In principle, this is a much more efficient use of radio telescope observations for determining the spatial distribution of \HI than traditional galaxy redshift surveys.  The \HI traces not only the galaxies, but also the underlying matter distribution, dominated by the dark matter. Observations of LSS enabled by 21\,cm intensity mapping can help us refine our models of cosmology and dark energy, and can shed light on the nature of dark matter. 


21\,cm IM was first proposed for studying \HI during cosmic dawn and the EoR to understand the formation of the first luminous objects  \citep{furlanetto_2006,pritchard_2012,2020PASP..132f2001L,shimabukuro_2023}. The techniques and challenges for these measurements are very similar to those of the post-EoR measurements, which are the focus of this paper. Note, however, that the lack of redshift catalogs of sources in the redshift range ($z\gtrsim 6$) limits the use of the cross-correlation techniques exploited by the first 21\,cm IM measurements at lower redshifts. IM is also performed with other bright spectral lines \citep{2022A&ARv..30....5B, 2026arXiv260203011C}, such as redshifted CO and [CII] lines at millimeter and sub-millimeter wavelengths and redshifted  H$\alpha$, H$\beta$, Ly$\alpha$, [OII], and [OIII] lines in the near-infrared \citep{HETDEX_2026,cheng_bayesian_2024}.
21\,cm IM is a subset of this broader field, which is generally called Line Intensity Mapping (LIM).


This review of 21\,cm intensity mapping begins, in \S\ref{sec:21cmIM}, with a discussion of specific LSS observables and the basic framework for intensity mapping.  In \S\ref{sec:instrumentconcepts} instrumentation that may be used to perform intensity mapping is described.  Particular attention is given to interferometers, the signal chain used to construct {\it visibilities}, which are the fundamental data output of these instruments, and the systematics associated with these instruments.  In \S\ref{sec:analysis} we describe how the data is processed to produce a useful representation of the radio intensity map as a function of angle and frequency and the challenging task of separating the 21\,cm signal from other radio emission contributing to this map. \S\ref{sec:obs21cmsignal} describes results that have been obtained by the intensity mapping technique while \S\ref{sec:future} describes future planned intensity mapping surveys.  The authors give their perspective on the progress made in 21\,cm intensity mapping and its future in \S\ref{sec:summary}.

\section{Post-EoR 21\,cm intensity mapping} 
\label{sec:21cmIM}

\subsection{Intensity Mapping, Large-Scale Structure and Baryon Acoustic Oscillations}

Traditional galaxy redshift surveys, which rely on detecting and spectroscopically confirming individual galaxies, are highly effective but are limited in volume and redshift coverage due to the significant telescope time required.  The alternative and complementary approach, 21\,cm IM measures the integrated 21\,cm line emission from neutral hydrogen (\HI) in the universe.  Each measurement contributes to the accumulated signal no matter the signal-to-noise, not having a threshold of detecting a galaxy.  Furthermore rather than resolving individual galaxies, 21\,cm IM measures the aggregate emission from all unresolved \HI sources within a given volume element. By mapping the \HI distribution over a large fraction of the sky and a broad redshift range (typically $z \sim 0.5-3$), 21\,cm IM surveys provides a dense, if noisy, three-dimensional map of the LSS in \HI.  The fluctuating brightness temperature of the 21\,cm signal then serves as a tracer of the underlying dark matter density field on large-scales.
 
The volumetric efficiency of this method allows for a vast number of independent modes to be sampled, leading to potentially unprecedented constraints on, for example, the BAO scale.  Baryon Acoustic Oscillation (BAO) are a key feature within the LSS. Originating from acoustic waves propagating in the primordial photon-baryon fluid before recombination, these oscillations imprinted a characteristic scale, the sound horizon at recombination, $r_s$. This scale manifests as a preferred clustering distance in the distribution of matter, observable as a prominent feature in the two-point correlation function or the power spectrum of density fluctuations. The BAO scale provides a standard ruler that can be measured at different redshifts. By comparing the observed angular scale, $\theta_{\rm BAO}(z)$, and the radial scale, $\Delta z_{\rm BAO}(z)$, to the known physical scale $r_s$, one can infer the angular diameter distance $d_A(z)$ and the Hubble expansion law $H(z)$. This expansion law is directly sensitive to the equation of state of dark energy, $w(z) = p_{\rm DE}/\rho_{\rm DE}$ which would be $-1$ only in the case of $\Lambda$ dark energy.  LSS via the BAO signal provides powerful constraints on dark energy and its evolution \citep{Bassett2010BAO} which can be improved by mapping LSS in larger cosmological volumes. Measurement of the BAO scale remain an important goal for optical redshift surveys 
\citep{Forero-Sanchez:2026zfb,Sanders:2026kgp}
and is a primary scientific driver for 21\,cm IM \citep{peterson200921cmintensitymapping, 2010ApJ...721..164S, 2015ApJ...803...21B} because it is one of the best probes of the cosmic expansion history and hence of the dark energy equation of state.


\subsection{The 21\,cm Power Spectrum and \HI  Properties}

The statistical properties of the 21\,cm brightness temperature field, $T_{21}(\mathbf{x})$, are typically quantified in Fourier space by the power spectrum $P_{\rm 21}(\mathbf{k})$, defined by
\begin{equation}
\langle \tilde{T}_{21}(\mathbf{k}) \tilde{T}_{21}^*(\mathbf{k}') \rangle = (2\pi)^3 \delta^D(\mathbf{k} - \mathbf{k}') P_{\rm 21}(|\mathbf{k}|),
\end{equation}
where $\tilde{T}_{21}(\mathbf{k})$ is the Fourier transform of the 21\,cm brightness temperature fluctuations. On large-scales relevant for BAO, the 21\,cm signal is expected to be a biased tracer of the underlying dark matter density field $\delta_m$. The relationship can be approximated by a deterministic bias model:
\begin{equation}
\delta T_{21}(\mathbf{k}) = \bar{T}_{21} (z) \left[ b_{\rm HI}(z) + f(z) \mu^2 \right] \delta_m(\mathbf{k}),
\end{equation}
where $\bar{T}_{21}(z)$ is the mean \HI brightness temperature, $b_{\rm HI}(z)$ is the \HI bias, $f(z)$ is the growth rate of structure, and $\mu = \hat{k} \cdot \hat{r}$ is the cosine of the angle between the wavevector and the line of sight, encoding the effects of redshift-space distortions (RSD). More generally one can always approximate the 21\,cm power spectrum as
\cite{1998ApJ...504..601P,1998ApJ...500L..79T,1999ApJ...520...24D}
\begin{equation}
P_{21}(k,z)\approx\bar{T}_{21}(z)^2\,\left[
(b_{\rm HI}(k,z)+f(z)\,\mu^2)^2+2\,(r_{21}(k,z)-1)\,b_{\rm HI}(k,z)\,f(z)\,\mu^2\right]\,
P_\mathrm{m}[k,z]
\end{equation}
in terms of a bias factor and correlation coefficient
\begin{equation}
b_{\rm HI}(|\mathbf{k}|,z)\equiv\frac{1}{\bar{T}_{21}(z)}\,
\sqrt{\frac{\langle|\delta T_{21}(\mathbf{k},z)|^2\rangle}
           {\langle|\delta_\mathrm{m}(\mathbf{k},z)|^2\rangle}}
\qquad
r_{21}(|\mathbf{k}|,z)\equiv
\frac{\langle\delta T_{21}(\mathbf{k},z)\,\delta_\mathrm{m}(\mathbf{k},z)^*\rangle}
    {\sqrt{\langle|\delta\,T_{21}(\mathbf{k},z)|^2\rangle\,
           \langle|\delta_\mathrm{m}(\mathbf{k},z)|^2\rangle}} \ .
\end{equation}
The non-deterministic bias represented by $r_{21}\ne1$ gives the stochasticity with which the \HI traces the matter distribution, similar to a stochasticity
affecting matter power spectrum estimates from galaxy surveys. 

The mean \HI density, parametrized by the cosmic \HI fraction $\Omega_{\rm HI}(z) = \rho_{\rm HI}(z)/\rho_{\rm crit,0}$, determines the overall amplitude of the signal through the relation \cite{Ibitoye2025}
\begin{equation}
\bar{T}_{21}(z) \approx 0.18 ~\mathrm{K}~ \Omega_{\rm HI}(z) h  \frac{(1+z)^2}{E(z)}
\end{equation}
where $E(z)=H(z)/H_0 = \sqrt{\Omega_m (1+z)^3 + \Omega_{\Lambda}}$.
Thus, measuring the 21\,cm power spectrum probes the abundance and clustering of \HI, characterized by $\Omega_{\rm HI} b_{\rm HI}$.
Note that for typical values of $\Omega_{\rm HI} \sim 10^{-3}$, the mean post-EoR brightness temperature would be $\bar{T}_{21} \sim 0.1\, \mathrm{mK}$, much lower than its expected value before reionisation \citep{pritchard_2012}. The signal weakness is however compensated by the lower foreground levels.  


\subsection{Post-Reionization 21\,cm Emission}

The nature of the 21\,cm signal changes drastically depending on the cosmic epoch. During and before the Epoch of Reionization (EoR), the signal is a complex mix of emission and absorption against the CMB. In the post-reionization era ($z \lesssim 6$), the intergalactic medium (IGM) is highly ionized, and the bulk of the neutral hydrogen resides within galaxies as atomic gas. In this regime, the spin temperature of the \HI, $T_s$, which governs the relative population of the hyperfine energy levels, is collisionally and radiatively coupled to the kinetic temperature of the gas. The gas in galaxies is heated by astrophysical processes such as star formation and active galactic nuclei, raising its temperature significantly above the CMB temperature, $T_s \gg T_{\rm CMB}$. Consequently, the 21\,cm signal is observed in emission against the cold CMB background, making it a relatively straightforward measure of the total \HI mass and its distribution. 

For an optically thin medium, the integrated 21\,cm intensity is proportional to the \HI column density and is generally independent of the spin temperature if $T_s \gg T_{\rm CMB}$ (the "high spin temperature" regime). In post-reionization galaxies, $T_s$ is typically higher than the CMB temperature. The connection between the \HI distribution (which traces the neutral gas in the ISM) and the underlying dark matter halos is governed by complex astrophysical processes, including galaxy formation, feedback, and the conversion of \HI to molecular hydrogen (H$_2$). These processes introduce the scale-dependent bias $b_{\rm HI}$ and a shot-noise term that must be modeled accurately for precision cosmology \citep{2017MNRAS.470.3220W, 2018ApJ...866..135V}.

\subsection{Foreground Contamination}

The greatest challenge for 21\,cm IM experiments is the formidable foreground contamination, which can be several orders of magnitude brighter than the cosmological signal. The main foreground components are:
\begin{itemize}
    \item \textbf{Galactic Synchrotron Emission:} This is the dominant foreground at 21\,cm frequencies ($\sim 100-1000$ MHz). It arises from relativistic electrons spiraling in the Galactic magnetic field. Its emission is characterized by a steep, featureless power-law spectrum in frequency, $T_{\rm synch} \propto \nu^{-\alpha}$, with a spectral index $\alpha \approx 2.5-2.7$. The emission is highly structured and anisotropic, correlating with the Galactic plane.
    \item \textbf{Galactic Free-Free Emission:} Produced by thermal Bremsstrahlung from ionized gas in the Milky Way (and other galaxies), this component has a slightly flatter spectrum, $\alpha \approx 2.1$, and traces the warm ionized medium (WIM).
    \item \textbf{Extragalactic Radio Sources:} These are divided into two categories: resolved point sources (e.g., active galactic nuclei), which can be identified and subtracted, and unresolved point sources (primarily faint radio galaxies and star-forming galaxies). The latter contribute a confusion noise that must be statistically modeled and accounted for. Their spectra are generally complex but often dominated by a steep power-law component.
\end{itemize}
A distinguishing characteristic of the 21\,cm signal is that it fluctuates in both frequency and angular position due to the clustering of \HI. Foregrounds, in contrast, are generally spectrally smooth across the relatively narrow frequency bands of an IM survey. This spectral smoothness is the primary lever arm for foreground removal techniques, which exploit the fact that the signal has significant structure in frequency (due to redshift-space fluctuations), while the foregrounds are dominated by low-order polynomials in frequency. A more detailed discussion can be found in section \ref{sec:fgndclean}.

\subsection{The Cylindrical Power Spectrum and Foreground Isolation}

The fundamental observable for 21\,cm IM is the three-dimensional power spectrum $P_{\rm 21}(\mathbf{k})$. Due to the cosmological principle, the statistical properties of the signal are isotropic. However, observations are made in a cylindrical volume, where we can distinguish modes perpendicular and parallel to the line of sight. The wavevector is decomposed into \mbox{$\mathbf{k} = k_\perp + k_\parallel$}, where $k_\perp$ corresponds to the angular structure on the sky and $k_\parallel$ corresponds to the frequency (or redshift) structure along the line of sight. The cylindrical or 2D power spectrum is then defined as $P(k_\perp, k_\parallel)$.

This separation is crucial for foreground mitigation. The spectrally smooth foregrounds are localized to low $k_\parallel$ modes (small frequency structure). In contrast, the cosmological signal spans a wide range of $k_\parallel$ values. Therefore, one can define a "wedge" in the $(k_\perp, k_\parallel)$ plane that is heavily contaminated by foregrounds. The boundary of this foreground wedge is determined by the horizon limit of the instrument, beyond which the signal is completely dominated by foregrounds \citep{2010ApJ...724..526D, 2012ApJ...752..137M}. This is further discussed in section \ref{sec:delayfgndwedge}. By discarding the modes within this wedge, the cosmological signal in the "EoR window" or "clean region" outside the wedge can be recovered, albeit at the cost of a loss of some information, particularly on large radial scales. The 2D power spectrum, $P_{21}(k_\perp, k_\parallel)$, and the foreground power spectrum, $P_{\rm fg}(k_\perp, k_\parallel)$, are thus the central theoretical constructs that guide the design and data analysis of 21\,cm IM experiments. The goal is to access the signal-dominated region of the $(k_\perp, k_\parallel)$ plane to extract the BAO signature and other cosmological information. 
 

\section{Instrument concepts} 
\label{sec:instrumentconcepts}

The information provided by the LSS is essentially statistical in nature, and one has to survey large volumes of the universe to characterize the matter distribution at different scales. In addition, observing the evolution of the LSS with the redshift is crucial to refine the cosmological model and study the nature of dark energy. 21\,cm  IM uses neutral atomic hydrogen (\HI) as a tracer of the underlying cosmic matter distribution, which is observed thanks to the redshifted \HI \, spin-flip transition at $\nu_{21} = 1420.4 \mathrm{MHz}$. IM surveys require thus radio instruments with a large instantaneous filed of view (FoV), to get the survey speed and a large observation frequency bandwidth to cover the redshift range. In this section, we outline the key characteristics of radio telescopes relevant to IM surveys. For a more comprehensive discussion, the reader is referred to \citep{2013pss2.book..283W, 2017isra.book.....T} and to \citep{2020PASP..132f2001L}. 

Because radio instruments operate at relatively long wavelengths, they are inherently diffraction-limited. A single-dish antenna with a diameter $D$ yields a primary beam angular size of $\theta_\mathrm{dish} \sim \lambda/D$, which spans $\theta_\mathrm{dish} \sim 3.5 - 35 \, \mathrm{arcmin}$ for a $D = 100\,\mathrm{m}$ aperture, and IM surveys with $\lambda \sim 10 - 100 \, \mathrm{cm}$. In non-interferometric (single-dish) observations, this primary beam dictates the on-sky angular resolution, $\delta \theta \sim \theta_\mathrm{dish}$. 

When constructing a 3D map, the transverse spatial resolution $\delta \ell_\perp$, expressed as a comoving length scale, is related to the angular resolution $\delta \theta$ via the comoving angular diameter distance $d_\mathrm{A}(z)$. Concurrently, the instrument's frequency resolution $\delta \nu$ defines its redshift resolution, mapping directly to the line-of-sight spatial resolution $\delta \ell_\parallel$ through the Hubble parameter $H(z)$ \citep[e.g.,][]{2012A&A...540A.129A, 2020PASP..132f2001L}. Together, these comoving structural resolutions establish the small-scale cutoff for accessible wavemodes\footnote{For simplicity of presentation we here use the small angle / small redshift range approximation where $k_\parallel$ and $k_\perp$ are approximated by Fourier transforms in angle-redshift space.}, $(k_\perp^{\max}, k_\parallel^{\max})$. Crucially, both comoving length scales exhibit a strong dependence on redshift ($z$), causing the effective spatial resolution to evolve significantly across cosmic time.
\begin{eqnarray}
    \delta \ell_\perp (z) & = & (1+z) \, \dang(z) \, \delta \theta \\
    \delta \ell_\parallel (z) & = & (1+z)^2 \frac{c}{H(z)} \frac{\delta \nu}{\nu_{21}} \\
    k_{\perp, \parallel}^{max} & \sim & \frac{2 \pi}{ \delta \ell_{\perp, \parallel} }
\end{eqnarray}

\begin{figure}[htpb]
\includegraphics[width=0.98\textwidth]{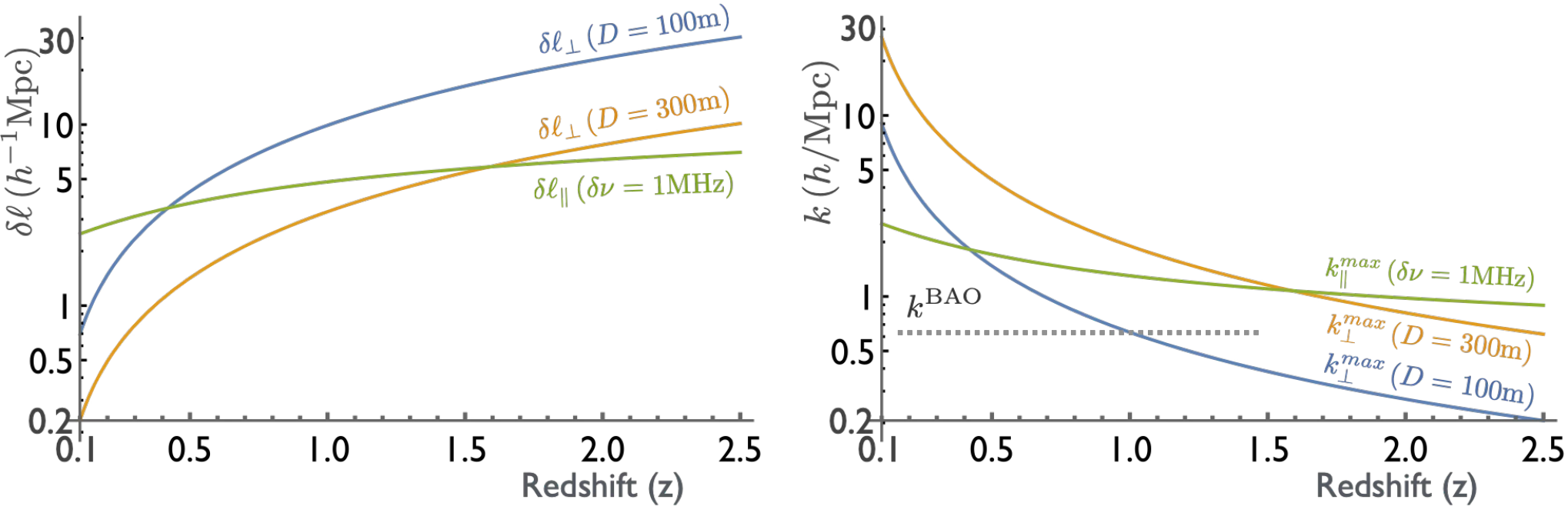} 
\caption{Left: The transverse length $\ell_\perp(z)$ resolution in $h^{-1} \mathrm{Mpc}$ as a function of redshift, for a $D=100, 300\, \mathrm{m}$ dish or array size (blue, orange), and the radial or line of sight resolution $\ell_\parallel(z)$, assuming $\delta \nu = 1 \,\mathrm{MHz}$ frequency resolution (green). Right: The corresponding maximum value of transverse $k_\perp^{max}$ and radial $k_\parallel^{max}$ wave numbers, in $h / \mathrm{Mpc}$. Also shown the approximate position of the first BAO wiggle $k^\mathrm{BAO} = 2 \pi / r_s$, where $r_s$ is the sound horizon at recombination,  assuming standard Planck 2018 \LCDM cosmology \citep{2021A&A...652C...4P}.  
\label{fig:comovlenresol}}
\end{figure}

Figure \ref{fig:comovlenresol} shows the variations of the transverse and radial resolutions,  assuming the Planck \LCDM model \citep{2021A&A...652C...4P}, for two dish sizes, $D=100, 300\, \mathrm{m}$ and $1 \,\mathrm{MHz}$ frequency resolution. The corresponding maximum  wave number values $k_{\perp, \parallel}^{max}$ are also shown. 

Another defining characteristic of radio astronomical instruments compared to optical telescopes and optical detectors is that their sensitivity is fundamentally constrained by receiver noise. Radio observations rely primarily on coherent detection, wherein incoming electromagnetic radiation is directly converted into a time-varying electric signal (voltage or current) and subsequently amplified using low-noise electronic amplifiers.

Noise levels in radio telescopes are conventionally parameterized by an equivalent temperature, known as the system temperature (\Tsys). This quantity represents the temperature of an equivalent blackbody sky that would produce a power spectral density identical to the total random noise generated by the receiver, its associated electronics, and environmental contributions. 

Despite substantial advancements in radio frequency (RF) technology, the noise temperature of the first-stage amplifier (e.g., a low-noise amplifier, or LNA) remains a primary contribution to the instrument noise budget, particularly within the UHF band near $1\,\mathrm{GHz}$. Additional contributions to \Tsys \, arise from stray light and ground spillover, where thermal radiation from non-sky sources enters the receiver via far side-lobes of the antenna primary beam \citep{2013pss2.book..283W}.

Diffuse sky emission, dominated by Galactic foregrounds, constitutes another major contribution to the overall system temperature. Neglecting minor atmospheric and CMB contributions, the total system noise budget \Tsys \, can be modeled as:
\begin{eqnarray}
    T_\mathrm{sys} & = & T_\mathrm{recv} + T_\mathrm{spill} + T_\mathrm{sky}
    \label{eq:tsys_budget}
\end{eqnarray}
where $T_\mathrm{recv}$ and $T_\mathrm{spill}$ denote the receiver noise temperature and ground spillover contribution, respectively, and $T_\mathrm{sky}$ represents the contribution to $T_\mathrm{sys}$ from atmospheric and astrophysical radiation, dominated by  
diffuse astrophysical foregrounds, particularly Galactic synchrotron emission.

Receiver noise temperatures $T_\mathrm{recv}$ of $50-70 \, \mathrm{K}$ are typical of carefully designed room temperature  wide band amplifiers, while $10-20 \, \mathrm{K}$ can be reached with cryogenically cooled first stage amplifiers \citep{2013pss2.book..283W}. 
The sky temperature at $408 \, \mathrm{MHz}$, dominated by the Galactic synchrotron emission, varies from $\sim 10 \, \mathrm{K}$ at high Galactic latitudes to more than $100 \,\mathrm{K}$ close to the Galactic plane \citep{1982A&AS...47....1H}. The synchrotron emission intensity varies with the frequency as $T_\mathrm{sync} \propto \nu^{-\beta}$ with $2.1 \lesssim \beta \lesssim 2.7$ \citep{2011A&A...525A.138G}. So, the synchrotron contribution to the system temperature is subdominant compared to receiver noise at high Galactic latitudes for the redshift range $z \lesssim 2.5$, however, it can become significant at the large redshift end of post-EoR 21\,cm IM surveys  $z \gtrsim 3.5$, reaching $T_\mathrm{gal} \sim 30-40 \,\mathrm{K}$ for the coldest parts of the sky. 

Sensitivity to faint sources is obtained thanks to large antennas, with collecting areas up to $\sim 10^4 - 10^5 \, \mathrm{m^2}$ for single dish radio telescopes\footnote{The Green Bank Telescope in the USA (\href{https://greenbankobservatory.org/about/telescopes/gbt/}{https://greenbankobservatory.org/about/telescopes/gbt/}) or the Effelsberg Radio Telescope in Germany (\href{https://www.mpifr-bonn.mpg.de/en/effelsberg}{https://www.mpifr-bonn.mpg.de/en/effelsberg}), with their 100 meter diameter dishes, are examples of fully steerable, currently operational large radio telescopes.
The FAST radio telescope in China, \href{https://fast.bao.ac.cn}{https://fast.bao.ac.cn}, with its huge 500 meter aperture fixed spherical reflector, is the largest single dish radio instrument.}.
Even with such large collecting area, long integration time is required to reduce noise fluctuations, and get signal to noise ratio above one for faint sources. Given a collecting area $A$, a total integration time $t_\mathrm{int}$ and a frequency band $\Delta \nu$, power collected from sources with a limiting flux density  (power per unit area per unit frequency, in $\mathrm{W/m^2/Hz)}$ or Jansky ($\mathrm{Jy}$)) larger than $S_\mathrm{lim}$ will exceed the noise average power, hence reaching a signal to noise ratio above one \citep{2017isra.book.....T, 2013pss2.book..283W}:
\begin{eqnarray}
\sigma^2_\mathrm{noise} = \frac{2 \mTsys^2}{t_\mathrm{int} \Delta \nu} & \rightarrow & 
    S_\mathrm{lim} = \frac{\sqrt{2} \, k_B \mTsys}{A \sqrt{t_\mathrm{int} \Delta \nu}},
\end{eqnarray}
where $k_B$ is the Boltzmann constant and we have assumed dual polarization observation. 
One obtains $S_\mathrm{lim} \sim 160 \, \mu\mathrm{Jy}$ for 1 hour integration time with a $A=10^4 \, \mathrm{m^2}, \mTsys=50\, \mathrm{K}$ telescope and $\Delta \nu=1\, \mathrm{MHz}$. This flux density limit $S_\mathrm{lim}$ corresponds to 21\,cm emission from a rather massive, Milky Way-like galaxy,  with $M_\mHI \sim 10^{10} \,\mMsol$ at a redshift $z \sim 0.25$.

\subsection{Single-dish observations}
Single-dish radio telescopes can be used to map the sky brightness by pointing the 
antenna toward different directions of the sky, measuring the brightness temperature as a function of frequency for each direction. A sky area is covered either through a set of 
pointed observations, followed by the telescope repointing (stare and step) or with the telescope scanning continuously over the target sky region.  A 3D sky map $T(\alpha, \delta, \nu)$ is then obtained by combining observations from different directions.
Consider a sky area $\Omega_\mathrm{survey}$ that will be observed with a $D$ diameter dish equipped with $N_f$ feed antennas, or $N_f$ single-feed identical dishes over the total duration $t_\mathrm{survey}^{tot}$. Assuming a solid angle resolution 
$\delta \omega = \left( \frac{\lambda}{D} \right)^2$, integration time $t_\mathrm{int}$ per resolution element can be written as: 
\begin{eqnarray}
    t_\mathrm{int} & = & N_f \, \left( \frac{(1+z) \lambda_{21}}{D} \right)^2 \left( \frac{1}{\Omega_\mathrm{survey}} \right) \, t_\mathrm{survey}^{tot}. 
    \label{eq:sd-surveytime}
\end{eqnarray}
The obtained brightness temperature maps $T(\alpha, \delta, \nu)$ with pixel size $\sim \delta \omega$ will have a noise fluctuation level $\sigma_T^2 = 2 \frac{2 \mTsys^2}{t_\mathrm{int} \delta \nu}$. Converting angles and frequencies to comoving 3D space coordinates, one gets a 21\,cm brightness map of a cosmological volume. The power spectrum 
$P_\mathrm{noise}( \vec{k} )$ of such a map will correspond to  white noise, spread over the parallelepiped volume $ k_\parallel^{max} (k_\perp^{max})^2$ in k-space. Expressed as brightness temperature fluctuations, the noise power spectral density $P_\mathrm{noise}( k_\perp, k_\parallel )= P_\mathrm{noise}^{SD}$ for single-dish IM surveys can be written as:
\begin{eqnarray}
    P_\mathrm{noise}^{SD} & \sim & 2 \frac{\mTsys^2}{t_\mathrm{int} \delta \nu} \frac{1}{\left( k_\perp^{max} \right)^2 k_\parallel^{max}} \\
    P_\mathrm{noise}^{SD} & \sim & \frac{2 \, \mTsys^2}{(2 \pi)^3 \, t_\mathrm{int} \nu_{21}} \left( \frac{\lambda_{21}}{D} \right)^2 \dang^2(z) \frac{c}{H(z)} (1+z)^6 \\
    P_\mathrm{noise}^{SD} & \sim & \left( 0.525 \, \frac{\mathrm{mK^2}}{\mathrm{Mpc^3}}\right) \left( \frac{\mTsys}{50 \mathrm{K}}\right)^2 \left( \frac{1 \mathrm{h}}{t_\mathrm{int}} \right) \left( \frac{100 \mathrm{m}}{D} \right)^2
    \left( \frac{\dang(z)}{10^3 \mathrm{Mpc}} \right)^2  \frac{1}{h_{70}(z)} (1+z)^6 
    \label{eq:sd-pnoise}
\end{eqnarray}
where $h_{70}(z) = H(z)/(70 \mathrm{km/s/Mpc})$. Note also the strong redshift dependence through the $(1+z)^6$ term, leading, for example, to a noise power of $P_\mathrm{noise}^{SD} \simeq 62.5 \mathrm{mK^2/Mpc^3}$ for a survey at  redshift $z=1$ with the above fiducial parameters $D=100\, \mathrm{m}, \mTsys=50 \,\mathrm{K} \ldots$. 
It should be noted that, although the noise power spectrum is flat, the sensitivity varies with $\vec{k}$, decreasing with $k_\perp$, as Fourier modes are damped for $k_\perp \gtrsim k_\perp^{max}(z)$ by the dish response function (antenna beam). 

As expected, a single dish instrument is rather inefficient to perform a sky survey, and two years of on-sky observations would be needed (eq. \ref{eq:sd-surveytime}) to survey $1000\, \mathrm{deg^2}$ at redshift $z=1$. 
One can increase the survey speed in single-dish IM surveys, parallelizing the sky observations by using multiple dishes or dishes equipped with Focal Plane Arrays (FPA), realized either by assembling multiple independent feed antennas at the focal plane, or through Phased Array Feeds (PAF), such as the Australian Square Kilometre Array Pathfinder (ASKAP) \citep{2021PASA...38....9H}. 
Using $N_d$ identical dishes, each equipped with a FPA with $N_f$ feeds, will decrease the total survey time $t_\mathrm{survey}^{tot}$ by a factor $N_d \times N_f$. 
The MeerKLASS \citep{2016mks..confE..32S} project implements the first strategy, using the 64 MeerKAT dishes \citep{2021MNRAS.502.2970A} to carry a 21\,cm IM survey (see section \ref{sec:meerkat}).

Traditional techniques used in single-dish radio observations, such as the ON/OFF method, to mitigate systematic effects caused by the variations in the instrument response, can not be used in IM surveys. 
The sky scanning strategy should be then carefully designed to mitigate such systematics effects, exploiting redundancies to overcome instrument response drifts and $1/f$ noise.  

\subsection{Interferometric observations}
The coherent electromagnetic signal detection in radio instruments makes interferometric measurements appealing and comparatively easy to implement, through combination of signals from pairs of antennas. This is especially useful to increase the angular resolution, illustrated by the record-holding angular resolution close to milliarcsec (mas), obtained with Very Long Baseline Interferometry (VLBI)\footnote{A history of space VLBI can be found in an arXiv paper by L. Gurvitz \href{https://arxiv.org/pdf/2306.17647}{https://arxiv.org/pdf/2306.17647}}
using baselines up to size of the Earth \citep{1999ASPC..180..433W}. 
We give here a brief overview of radio interferometry, focusing on sensitivity for 21\,cm IM. Detailed presentation of the subject can be found in \citep{2017isra.book.....T} or \citep{2013pss2.book..283W, 2013pss2.book..325M}.

We consider a pair of identical antennas $(i,j)$, coplanar with respect to their common axis, pointed toward a given sky direction.
The correlation of the two RF signals, called  the visibility ${\cal V}_{ij}(\nu)$, is obtained by the time average of the complex amplitudes $(F_{i,j}(\nu))$ of the two signals in a narrow band around the frequency $\nu$ :  
\begin{eqnarray}
    {\cal V}_{ij}(\nu) & = & \langle F_i(\nu) F_j^*(\nu) \rangle_\mathrm{time},
\end{eqnarray}
where $F^*$ denotes the complex conjugate of $F$.  For a narrow primary beam corresponding to a small FoV, 
we write the sky brightness temperature $T(\xi, \eta, \nu)$ as a function of the frequency $\nu$ and two angular coordinates $(\xi, \eta)$, centered on the antennas pointing direction, in the flat-sky approximation.
The sky brightness ${\cal T}(u,v,\nu)$ and the antenna primary beam ${\cal L}(u,v, \nu)$ are represented in the Fourier plane as a 
function of the two angular frequencies $(u,v)$, Fourier conjugate 
variables of the two angles $(\xi, \eta)$ over the sky plane. 
\begin{eqnarray}
    {\cal V}_{ij}(\nu) & = & \iint \, {\cal T}(u,v,\nu) {\cal L}(u - \frac{\Delta x}{\lambda}, v - \frac{\Delta y}{\lambda}, \nu) \, du \, dv  \label{eq:visisky}\\
    \vec{b}_{ij} & = & (x_j - x_i, y_j - y_i) = (\Delta x, \Delta y)) \\
    T(\xi, \eta, \nu) & = &  \iint \, {\cal T}(u,v,\nu) \, \exp \left( 2 i \pi ( u \xi + v \eta )  \right) \, du \, dv .
\end{eqnarray}
The visibility ${\cal V}_{ij}$ is a measure of the complex Fourier coefficient of the sky intensity $I(\xi, \eta, \nu)$, for the angular frequency $\left(u_{ij} , v_{ij} \right)$, defined by the  baseline vector  $\vec{b}_{ij} = \vec{r}_j - \vec{r}_i$:  
\begin{eqnarray}
\left(u_{ij} , v_{ij} \right) & = & \frac{\vec{b}_{ij}}{\lambda} = \vec{b}_{ij}\frac{\nu}{c} = \left( \frac{\Delta x}{\lambda},  \frac{\Delta y}{\lambda} \right),
\end{eqnarray}
where the $x,y$ axes are parallel to the two angular planes $(\xi, \eta)$. More precisely, ${\cal V}_{ij}$ is an average of Fourier coefficients in a region of the angular frequency plane $(u,v)$ around $\left(u_{ij} , v_{ij} \right)$, weighted by the 
antenna beam ${\cal L}(u,v, \nu)$. If the antenna is a $D$ diameter dish, the $(u,v)$ plane region would be a disk of radius $ D / \lambda$.
When the flat-sky approximation is invalid, such as for interferometers with a 
large FoV or when mapping wide celestial areas, spherical harmonics must be employed to represent both the sky intensity and the instrument's 
beam response. 
Consequently, the visibility measured by a given baseline $\vec{b}_{ij}$ 
corresponds to a weighted sum of the sky's spherical harmonic coefficients, $\mathcal{T}(\ell, m, \nu)$. This summation occurs over a region in the 
$(\ell, m)$-plane corresponding to the antenna pair beam response 
$\mathcal{L}_{ij}(\ell, m, \nu)$, which itself is determined by the baseline, the dish primary beam, and the pointing direction \citep{2014ApJ...781...57S, 2016MNRAS.461.1950Z}. The extension to the polarised sky case, not discussed here, can be found in \citep{2015PhRvD..91h3514S}.

Radio interferometry has traditionally delivered high angular resolution by utilizing long baselines $(|\vec{b}_{ij}| \gg D)$ relative to the individual antenna diameter $D$. With a few baselines scattered over a vast geographic footprint, they naturally yield a sparse sampling of the $(u,v)$ plane. To mitigate this limitation, observers rely on rotation synthesis \citep{2017isra.book.....T, 2013pss2.book..325M}. In this operational mode, all antennas track a fixed target in the sky, continuously compensating for the Earth's rotation. As tracking progresses, the projected baselines on the sky dynamically shift, with each baseline tracing out an arc in the plane perpendicular to the target direction. While this rotation synthesis strategy dramatically improves $(u,v)$ coverage, it demands a significant investment of time, typically requiring at least several hours of continuous observation per target field.

\subsection{Packed array transit interferometers}
As shown above in section \ref{sec:21cmIM}, the statistical properties of the matter distribution at large-scales (LSS), from the largest angles on the sky up to a fraction of a degree ($\sim 10 \mathrm{arcmin}$), contain valuable information for testing the cosmological model and constrain the dark energy equation of state, especially thanks to the BAO features. However, this information being statistical, a large cosmic volume, and hence a large area of the sky, needs to be surveyed to measure the LSS properties with low enough statistical uncertainties. Typical existing interferometer configurations, which aim at getting high angular resolution over small sky areas, are thus not well-adapted for IM surveys. 

As an alternative to large dishes, 100 meter size or above and equipped with a FPA, 
densely packed interferometric arrays utilizing relatively small reflectors, 5 to 10 meter size and operating in transit mode, have been proposed as the optimal architecture for a 21\, cm IM instrument. A few dedicated packed array interferometers have been built, such as CHIME \citep{2022ApJS..261...29C} and Tianlai \citep{2015ApJ...798...40X}, and others are planned or being built (see section \ref{sec:future}).

The small reflector diameter ($D \lesssim 10 \mathrm{m}$) ensures a large instantaneous field of view with \mbox{$\mathrm{FoV} \sim \left( (1+z) \lambda_{21} / D \right)^2 \,\mathrm{sr}$}, while the dense configuration concentrates the survey sensitivity within the $k_\perp$ range most relevant to probing LSS. Interferometers operating in transit mode have all their reflectors pointed toward a given sky declination $\delta_o$, in the meridian plane. Observation is carried out in drift-scan mode and a sky stripe, centered on $\delta_o$, with a declination width given by the reflector's primary beam, passes through the instrument beam thanks to the Earth's rotation. A map of the sky stripe, over the full 360 degree right ascension range can be obtained after a complete Earth rotation, i.e. 24 hours. The integration time, and hence the sensitivity, is increased by observing the same sky stripe many times, over few months or even years. The sky coverage can be increased by observing at several declinations. 
Transit-mode operation is ideally suited for mapping extensive areas of the sky while minimizing instrument complexity and manufacturing costs. In addition, having no moving parts or instrument configuration changes during observations helps improve instrument stability, easing, in principle, the calibration.  

If the observed sky 21\,cm brightness temperature $T_{21}(\xi, \eta, \nu)$ is projected on a universe shell at redshift $z$, with $1+z = \frac{\nu_{21}}{\nu}$, and the angles are converted to comoving transverse distances, then the $(u,v)$ angular frequencies of the baseline would be mapped to a transverse wavenumber $k_\perp$ through the relation:
\begin{eqnarray}
     \vec{k}_\perp(z) & = & \frac{2 \pi}{(1+z)^2 \dang(z)} \frac{\nu_{21}}{c}  \vec{b}_{ij} \simeq \left(3 \, \mathrm{Mpc}^{-1}\right) \left( \frac{1000 \, \mathrm{Mpc}}{(1+z)^2 \, \dang(z) } \right) \left( \frac{\vec{b_{ij}}}{100 \,\mathrm{m}} \right).
\end{eqnarray}

Assuming uncorrelated noise fluctuations to the RF signal of different antennas, the  noise contribution to the visibility averages out to zero $\langle {\cal{V}}_{ij}^{noise} \rangle = 0$. This is a significant advantage of interferometric measurements, compared to single dish observations, where, after detection, the noise produces an offset which is much larger than the signal. Visibilities would then be an unbiased measurement of the sky signal power spectral density, rather insensitive to the receiver temperature variations and $1/f$ noise. 
However, some of the contributions to the system temperature might give rise to a correlated component in the noise, which will then show up as an offset or bias in the measured visibility. 

Noise fluctuations induce a dispersion in visibility measurements with an RMS value of $\sigma_{\cal V}^2 \sim 2 \mTsys^2 / \left( t_\mathrm{int} \delta \nu \right)$.  
As a visibility represents a measurement of the sky Fourier 
modes sampled over a specific $(u,v)$-plane resolution element
$ \delta u \times \delta v \sim \left(D / \lambda \right)^2 $, 
the associated measurement noise can represented as a noise power spectral density (PSD) in the $(u, v)$ plane:
\begin{eqnarray}
    P_\mathrm{noise}^{\cal V}(u_{ij}, v_{ij}) & \simeq & \frac{2 \mTsys}{t_\mathrm{int} \delta \nu} \left( \frac{1}{\delta u \times \delta v} \right) \sim \frac{2 \mTsys}{t_\mathrm{int} \delta \nu} \left( \frac{\lambda}{D} \right)^2 \label{eq:interfnoiseuv}
\end{eqnarray}
Converting $(u,v)$ to wave mode $k_\perp$, and the frequency $\nu$ to the redshift and $\delta \nu$ to $k_\parallel^{max}$, one can obtain the expression noise PSD $P_\mathrm{noise}^{\cal V}(\vec{k})$ in k-space for each visibility measurement, 
similar to the single-dish case (eq. \ref{eq:sd-pnoise}). Note that $D$ in the above expression corresponds to the interferometer individual reflector diameter, and not the array size. 
In densely packed arrays, a large number of nearly identical short baselines 
are formed. This inherent redundancy significantly boosts survey sensitivity 
at small transverse wavenumbers ($k_\perp$) corresponding to these short baselines.
Specifically, the total noise PSD of the 
interferometer at these scales is reduced, as the single-baseline noise PSD 
is divided by $N_b$, the number of redundant baselines.

An array of cylindrical reflectors with their longitudinal axes aligned 
North-South were thought to represent a near-optimal configuration for IM transit interferometers \citep{2006astro.ph..6104P}. 
Each cylinder is equipped with closely packed receivers and feed antennas along its axis. Because each cylinder focuses radiation exclusively in the 
East-West direction, each receiver sees an elongated patch of the sky 
centered on the meridian; this FoV is narrow in the East-West direction but spans widely from horizon to horizon along the North-South direction. 
For a 10-meter-wide cylinder, the instantaneous FoV is exceptionally large, exceeding $\sim 100\text{ deg}^2$. Furthermore, for a half-wavelength ($\lambda/2$) feed spacing, all sky directions within the 
meridian plane can theoretically be reconstructed with an angular resolution determined by the instrumented length of the cylinders. For regularly spaced feeds, this can be performed more efficiently using a Fast Fourier Transform (FFT)-inspired beamforming. \citet{2009PhRvD..79h3530T} extended this idea to a 2D array of regularly spaced antennas or Omniscope \citep{2010PhRvD..82j3501T}.  

The CHIME \citep{2022ApJS..261...29C} and Tianlai \citep{2020SCPMA..6329862L} experiments have constructed cylindrical radio telescopes (CRTs) and are currently conducting surveys with these instruments. 
However, data processing and analysis have proven more challenging than anticipated. This difficulty is primarily driven by significant electromagnetic cross-coupling between closely spaced feeds along a given cylinder. 
Cross-coupling between feeds in densely packed dish arrays has also been 
identified as a primary limiting factor for these instruments \citep{2024JAI....1350002K, 2019ApJ...884..105K, 2024MNRAS.534.2653M}. To mitigate this issue, carefully designed reflector feed systems with low sidelobe levels are essential to maintain cross-coupling at an acceptable level. 
\subsection{Digital interferometry}
To detect LSS, interferometric instruments for 21\,cm IM surveys require wide frequency bands (hundreds of megahertz) and a massive number of elements, often exceeding a thousand antennas, to achieve the necessary sensitivity and survey speed. This scale would be nearly impossible to manage without the development of digital interferometers, which process the majority of the signals in digital systems using a combination of custom FPGA-based electronics and CPU/GPU clusters. We briefly describe the primary system components below; more detailed technical overviews can be found in \citep{2020PASP..132f2001L}, as well as in \citep{2022ApJS..261...29C, 2020SCPMA..6329862L} for specific instruments.

The signal path begins with individual RF signals from 
each feed and polarization, which pass through an analog amplification 
and bandpass filtering chain. If necessary, a mixer and local oscillator 
are used to down-convert the frequency \citep{2013pss2.book..283W} before 
digitization by an analog-to-digital converter (ADC). To satisfy the 
Nyquist theorem, the sampling frequency $f_s$ must be at least twice the 
total bandwidth, $\Delta \nu = \nu_{\max} - \nu_{\min}$, yielding typical 
sampling rates of $f_s$ of a few hundred MHz. The 
subsequent {\it F-engine} stage decomposes the digitized RF signal into its 
frequency components. This spectral decomposition is typically performed 
using a windowed Fourier transform, applying an FFT to blocks of $N$ consecutive waveform samples. The sequence of time samples can be apodized to minimize spectral leakage and reduce sidelobe contamination. This yields $N/2$ frequency channels covering the range $\nu_{\min} < \nu < \nu_{\max}$ with a channel resolution of $\delta \nu = f_s / N$. 
The aggregate input data rate to an F-engine processing $N_{RF}$ signals, sampled at the frequency $f_s$ scales as $N_{RF} \times f_s$, corresponding to $\sim 1\,\mathrm{TB/s}$ 
for $f_s = 1\,\mathrm{GHz}$ with 8-bit samples and $N_{RF} = 10^3$.

Processing time-stream blocks with a typical value of $N = 8192$ samples yields 4096 spectral 
channels, providing a frequency resolution of $\delta \nu \approx 122\,\mathrm{kHz}$ 
with a time resolution of $\approx 10\,\mu\mathrm{s}$. These 
Fourier components, or channelized data $F_i(\nu)$, are then routed to the 
\textit{X-engine}. The X-engine computes the cross-correlations between 
all signal pairs (visibilities) by averaging over an integration time window, yielding a time stream of frequency-dependent visibilities for each baseline:
\begin{align}
    \mathcal{V}_{ij}(t, \nu) &= \langle F_i(\nu) F^*_j (\nu) \rangle_{\mathrm{time}} 
    = \frac{1}{M} \sum_{m=1}^M F_i(\nu, t_m) F^*_j (\nu, t_m).
\end{align}
The transfer of channelized data from the F-engine to the X-engine presents 
a major technical challenge, as it requires a massive data transpose 
operation (often referred to as a ``corner turner''). At the F-engine output, 
the data are grouped by antenna, containing all frequency channels for a 
given RF signal. In contrast, the X-engine requires the data be grouped 
by frequency, pooling the Fourier components of all antennas for a single 
frequency channel to compute the cross-correlations.

For a transit interferometer with maximum baselines of a few hundred meters, 
a visibility integration time of approximately one second is well-matched 
to the instrument's angular resolution of a few arcminutes and the rate of sky rotation. Assuming an 
array of $\mathcal{O}(10^3)$ elements with the aforementioned sampling parameters, each visibility measurement averages $M \approx 10^5$ individual 
Fourier components. This requires an X-engine computational throughput of 
approximately $10^{15}$ complex multiply-accumulate (CMAC) operations per second, scaling as $\propto N_{RF}^2 f_s$.

\subsection{Systematic effects}

All 21\,cm  IM measurements to date are limited by systematic effects rather than raw sensitivity or integration time.  The most important effects are those that introduce a frequency-dependent response of the instrument, which convert bright, smooth-spectrum foregrounds into frequency-dependent contaminants that masquerade as the \HI signal. 
The foremost of these effects is frequency-dependence (or ``chromaticity") in the beam patterns of the antennas.  Because the foregrounds are $\sim 10^5$ times brighter than the 21\,cm signal, the patterns must be known to high precision.

For interferometers, a key challenge is knowing the primary beam patterns of the antennas in the array \citep{2024MNRAS.534.2475C}.  Each visibility involves integrating over the product of the primary beam patterns of the two antennas that define the baseline times a phase factor corresponding to the optical path difference between them:  
$\exp(i \vec{k}_{EM} \cdot \vec{b}_{ij})$, where $|\vec{k}_{EM} | = 2 \pi / \lambda$.  The phase factor is defined by the locations of the antennas, which can be measured quite accurately.  The beam patterns are more difficult to measure because what's needed is the beam pattern of each element {\it when embedded in the array}. These patterns are different from the patterns measured for isolated antennas because they interact with each other through ``mutual coupling" -- radio waves from the sky scatter between the antennas in the array and distort the antenna patterns.  This effect is accentuated in arrays designed for IM, which favors dense arrays, with small antenna separation compared to their size (i.e.$| \vec{b}_{ij} | \gtrsim D $). The latter enables good coverage of the $(u,v)$ or $k_\perp$ plane.  The coupling effects are complicated so they must either be modeled accurately or measured in situ. Once the full patterns are known, including effects of mutual coupling, corrections can be applied to recover the underlying 21\,cm power spectrum \citep{2019ApJ...884..105K, 2024MNRAS.534.2653M, ohara_direct_2026}.  

Multiple approaches have been pursued for determining the primary beam patterns of interferometers.  Since the antennas in drift-scanned transit interferometers generally can’t be pointed easily, they can’t be scanned across bright radio sources to measure the patterns.  Electromagnetic simulations are now capable of modeling even large arrays \cite{2022RAA....22f5020S}, but modeling all the details of an actual, “as-built” array is likely impossible at the required level of accuracy.  Promising results have come from drone-mounted radio transmitters flown above antenna arrays\citep{2017PASP..129c5002J,2026RAA....26b5001L}.  Drone measurements of CHIME \citep{tyndall_beam_2025} agree at least qualitatively with other methods for probing portions of the primary beams, using the Sun or holography, and the drone measurement of Tianlai also yields results roughly in agreement with measurement using the sky model, though there are differences in quantitative details \cite{2026RAA....26i5024G}.  In cases such as CHIME, where the far-field radiation pattern is at altitudes beyond the reach of drones, measurements of the beams in the near-field can be performed, with the plan of transforming them to the far-field.  Finally, serendipitous use of radiation emitted from satellites (GNSS/GPS) has been exploited in frequency bands that overlap the 21,cm IM bands, such as for CHORD\citep{2026PASA...43...16B} and PAON4\citep{2020MNRAS.493.2965A}.

Single-dish instruments (GBT, Parkes, FAST, MeerKAT, etc.)  have simpler beam patterns - a significant advantage for this approach compared to interferometers.   However, achieving adequate sensitivity requires having many such dishes with single-feed antennas and receivers (i.e. MeerKAT/SKA-Mid) or a single dish with an FPA \cite{1996PASA...13..243S}.  
The packed array antennas may still have mutual coupling between the antennas \cite{2024JAI....1350002K} which affects the beam \cite{2021IAPM...63f..98Z}, while
the latter may have mutual coupling between antennas in the focal plane. Some of these instruments are capable of scanning, so the patterns could be measured on bright radio sources. Primary beam patterns of MeerKAT antennas have been determined using a combination of astroholography (cross-correlating measured voltages from an antenna under test with those from another antenna tracking the same
source) and electromagnetic simulation\citep{2021MNRAS.502.2970A}. In any case, the frequency-dependent antenna patterns need to be measured (and remeasured in case they change with time).   


\section{Data Analysis challenges} 
\label{sec:analysis}
This section outlines the main processing stages required to extract the cosmological 21\,cm signal from radio observations, focusing specifically on the power spectrum $P_{21}(\vec{k})$ measurement. For a more in-depth treatment of these methodologies, we refer the reader to \citep{2013pss2.book..283W, 2013pss2.book..325M, 2019MNRAS.483.2207M} or the comprehensive review by \citet{2020PASP..132f2001L}. 

Raw data from a radio telescope array are typically structured as 3D arrays, where the first two dimensions correspond to time and frequency, and the third one to a specific baseline or element index. For a single-dish telescope equipped with a multi-feed receiver, or for an array of independent dishes, this third dimension maps to each individual feed or antenna pair. The initial stages of the data pipeline analyze these 2D time-frequency datasets per feed or baseline to detect and flag Radio Frequency Interference (RFI), as well as to mitigate instrumental artifacts such as baseline drifts and $1/f$ noise.

RFI manifests in diverse forms. The most prevalent types include persistent, statistically significant power excesses confined to specific frequency channels, as well as transient, broadband power bursts that simultaneously impact multiple channels. A review of real-time RFI mitigation techniques is provided by \citet{2025arXiv251201954B}, while \citet{2024MNRAS.529.4719G} demonstrates the utility of a Convolutional Neural Network (CNN) specifically tailored for automated RFI detection and flagging.

\subsection{Calibration}
The calibration process for radio interferometers requires determining three core sets of parameters: the passband response (the relative response across different frequencies), the absolute amplitude scale (converting measured electrical signal strengths to physical sky brightness temperatures), and the relative amplitudes and phases of the individual array elements. The measured visibilities $\mathcal{V}_{ij}^{\text{meas}}(\nu)$ can be modeled as a function of the true sky visibilities via:
\begin{equation}
    \mathcal{V}_{ij}^{\text{meas}}(\nu) = g_i(\nu) g_j^*(\nu) \, \mathcal{V}_{ij}(\nu) + n_{ij}  \label{eq:visicomplxgains}
\end{equation}
where $g_i(\nu)$ and $g_j(\nu)$ denote the frequency-dependent complex gains (capturing both amplitude and phase instrumental effects) for elements $i$ and $j$, respectively, and $n_{ij}$ represents the noise contribution. 
Furthermore, the true visibility is linked to the underlying sky brightness distribution through the primary beam pattern and the baseline configuration (Equation~\ref{eq:visisky}). Refining the primary beam model is therefore often an integral part of the overall calibration pipeline. 
Note that Equation~\ref{eq:visicomplxgains} must be formulated in a polarization-dependent manner. While an ideal instrument with orthogonal polarization probes in each feed can be calibrated independently across polarizations, real-world instruments suffer from polarization leakage. This couples baselines of different polarization and introduces non-zero signals into the cross-polarization visibilities, which need to be included for calibration. 

The first calibration method is sky-based and applicable to all interferometers. Starting with a model of the radio sky and the primary beam response, a set of expected true visibilities is computed. These visibilities then serve as input to a standard optimization algorithm, such as maximum likelihood, to solve for the complex gains.

For instruments with narrow primary beams, sky-based calibration is often achieved by targeting a bright, unresolved point source whose flux dominates the measured visibilities. In contrast, instruments with wide primary beams require a reasonably complete sky model and primary beam response. Such wide-field instruments often consist of many antennas with numerous identical baselines, making them well-suited for redundant baseline calibration \citep{2010MNRAS.408.1029L}. Indeed, with a sufficiently high degree of redundancy, it is possible to determine both the complex gains and the true visibilities, which are shared unknowns across all antenna pairs with identical baselines. However, fundamental degeneracies remain, leaving the absolute amplitude scale and an overall phase gradient undetermined. While hybrid calibration \citep{2018ApJ...863..170L} attempts to combine sky-based and redundant approaches, its precision is limited by sky model incompleteness, rendering it insufficient to detect the faint cosmological 21\,cm signal from the Epoch of Reionization (EoR) \citep{2016MNRAS.461.3135B, 2018AJ....156..285J, 2019ApJ...875...70B}. Crucially, because the bias induced by calibration errors on the extracted signal is roughly proportional to the bright foreground emissions, the calibration requirements for intensity mapping surveys 
are highly demanding.  

\subsection{Map making}
Reconstructing the sky brightness distribution $T(\alpha, \delta, \nu) \equiv T_p(\nu)$ from observed visibilities (or autocorrelation signals in the case of single-dish surveys) $V_{ij}(\nu, t_k) \equiv {\cal V}_{ij,k}(\nu)$ can be framed as solving a linear inverse problem, treated often independently for each frequency, 
\begin{equation}
    {\cal V}_{ij,k} = \mathbf{B} \, T_p + n   \label{eq:mapmaking}
\end{equation}
where ${\cal V}_{ij,k}$ is a column vector gathering visibilities from all baselines with $k$ indexing the time sample, $T_p$ is a vector representing the sky brightness with $p$ corresponding to the sky direction $(\alpha, \delta)$ pixel index, and $n$ denotes the noise vector. The operator matrix $\mathbf{B}$ encodes both the scanning strategy and the convolution by the angular response of the antenna or baseline beams. 

This formulation represents a classic inverse problem, extensively developed within the CMB community over several decades \citep[e.g.,][]{1997ApJ...480L..87T}. While easily generalized to polarized emission and typically solved independently for each radio frequency $\nu$, map-making plays distinct roles depending on the instrument geometry: it is essential for single-dish experiments prior to foreground removal and power spectrum estimation, whereas interferometric pipelines can often bypass it entirely. A detailed evaluation of map-making specifically tailored to 21\,cm intensity mapping, including the derivation of sky map 
and power spectrum uncertainties and associated covariance matrices, is presented by \citet{2015PhRvD..91b3002D}.

Reconstructing the sky brightness vector from visibilities poses two major challenges beyond the need for calibrated data and precise primary beam models. First, the problem is computationally demanding due to the massive scale of the matrix $\mathbf{B}$, which typically contains $10^6 - 10^8$ rows and $10^5 - 10^7$ columns. Second, Equation~\ref{eq:mapmaking} is inherently under-constrained owing to incomplete sampling of the spatial frequency $(u,v)$ plane. A linear solution can be obtained by applying a regularization scheme, or one can employ non-linear iterative methods such as CLEAN \citep{1974A&AS...15..417H}. The CLEAN algorithm and its subsequent refinements \citep[e.g.,][]{2008ISTSP...2..793C} model the sky as a collection of point sources by iteratively identifying the brightest peak in the dirty image and subtracting its contribution from the visibilities.

For transit interferometers that observe the full $360^\circ$ of the sky along the Right Ascension (RA) direction, one can perform a Fourier transform in RA, known as the $m$-mode transform. The maximum $m$-mode ($m_{\text{max}}$) is determined by the angular resolution of the visibilities, hence the corresponding averaging time. Consequently, the full linear map-making system in Equation~\ref{eq:mapmaking} decouples into $m_{\text{max}}$ independent, significantly smaller subsystems. This drastically eases the computational burden, as the individual $m$-modes can be processed independently in parallel \citep{2014ApJ...781...57S, 2015PhRvD..91h3514S, 2016MNRAS.461.1950Z, 2016RAA....16..158Z}.

Rapid advancements in artificial intelligence have led to the emergence of nonlinear image reconstruction techniques based on neural networks and machine learning, with \citet{2023A&A...677A.167G} providing one such example for interferometric map-making. Alternatively, sky maps can be reconstructed using closure quantities \citep{1974ApJ...193..293R, 2017isra.book.....T} formed from combinations of three or four visibilities. Map-making methods built on closure relations \citep[e.g.,][]{2018ApJ...857...23C} are intrinsically immune to antenna-based calibration errors. For instance, the phase of the bispectrum product, formed by three visibilities ${\cal V}_{ij} {\cal V}_{jk} {\cal V}_{ki}$ along a closed baseline triangle, is entirely independent of the individual antenna phase errors.

\subsection{Foreground removal}
\label{sec:fgndclean}
As discussed above, Galactic and extragalactic foregrounds are $\sim 10^4 - 10^5$ times brighter than the cosmological neutral hydrogen signal at redshifts $z \lesssim 3$ \citep[e.g.,][]{2014MNRAS.444.3183A, 2019arXiv190912369C}. Despite this extreme dynamic range, previous studies have demonstrated that foreground subtraction can preserve the 21\,cm signal at sufficiently large $(k_\perp, k_\parallel)$ values \citep[e.g.,][]{2012A&A...540A.129A, 2020MNRAS.499..304C}. Although numerous foreground removal algorithms have been proposed, they all exploit the spectral smoothness of foreground emissions \citep[e.g.,][]{2015MNRAS.447..400A, 2019arXiv190912369C}, which naturally arises from the underlying power-law electron energy distribution driving synchrotron radiation \citep{2021A&A...651A.116P}.

The reconstructed sky brightness temperature $T(p, \nu)$ in a given sky direction (or pixel $p$) at frequency $\nu$ can be expressed as the linear sum of three distinct contributions:
\begin{equation}
    T(p, \nu) = T_\mathrm{fg}(p, \nu) + T_{21}(p, \nu) + T_\mathrm{noise}(p, \nu)
\end{equation}
where $T_\mathrm{fg}$ represents the astrophysical foregrounds, $T_{21}$ is the target cosmological 21\,cm signal, and $T_\mathrm{noise}$ accounts for instrumental noise. Furthermore, we assume that the total foreground signal can be decomposed into $C$ distinct components, each defined by a spatial amplitude $a_c(p)$ and a frequency-dependent spectral profile $w_c(\nu)$:
\begin{equation}
    T_\mathrm{fg}(p, \nu) = \sum_{c=1}^{C} a_c(p) w_c(\nu).
\end{equation}
In the simplest parametric foreground removal approach, each spectral component $w_c(\nu)$ is modeled by a parameterized functional form characterized by a set of shape parameters $\{h_c^q, q=1 \dots Q_c\}$. The spatial amplitudes $a_c(p)$ and local shape parameters $h_c^q(p)$ are fitted independently along each pixel's line of sight $p$. Subtracting these best-fit foreground profiles from $T(p, \nu)$ yields a residual map containing, in principle, only the cosmological signal and instrumental noise.

In blind foreground separation methods, the foreground component shapes $w_c(\nu)$ are unknown, but assumed to be the same for all sky pixels. The shapes are then determined using the statistical properties of the observed 3D sky maps $T(p, \nu)$, and an a priori hypothesis concerning the foreground component. 

Principal Component Analysis (PCA)  is a widely used and effective method for reducing the dimensionality of large datasets \citep{2016RSPTA.37450202J}. It has been successfully applied to IM surveys \citep[e.g.,][]{2019AJ....157....4Z} to extract the spectral components $w_c(\nu)$, assuming that the spatial amplitudes $a_c(p)$ are uncorrelated, normally distributed random variables. These principal components are obtained by diagonalizing the observed frequency-frequency covariance matrix:
\begin{equation}
    \mathrm{Cov}(\nu, \nu') = \langle T(p, \nu) T(p, \nu') \rangle_p.
\end{equation}
Independent Component Analysis (ICA) can be viewed as a higher-order extension of PCA to non-Gaussian distributions, requiring only statistical independence among the underlying components \citep{1994SigPr..36..287C}. Alternative blind source separation methods that exploit signal sparsity in a transform domain, such as Generalized Morphological Component Analysis (GMCA), have also been adapted for 21\,cm foreground removal \citep{2020MNRAS.499..304C, 2026A&A...706A...3G}. More recently, deep learning (DL) architectures evaluated on simulated datasets \citep{2021JCAP...04..081M, 2024MNRAS.532.2615C, 2024PhRvD.109f3509S} have demonstrated superior performance compared to traditional linear techniques. Broad overviews of deep learning accessible to non-specialists can be found in \citep[e.g.][]{2015Natur.521..436L, 2015NN.....61...85S, 2019ieee53040, 2021jbd1186}, while a review focusing on astrophysical applications is provided by \citet{2026annurev10.1146}.

To date, 21\,cm signal extraction efforts on observational data have primarily relied on PCA, or ICA in select cases, such as analysis of Green Bank Telescope (GBT) IM data \citep{2017MNRAS.464.4938W}. Although PCA offers the practical advantage of absorbing residual bandpass calibration errors, it suffers from two notable limitations: selecting the optimal number of modes to subtract lacks a rigorous a priori criterion, and the resulting physical interpretation of the removed modes remains challenging.

\subsection{Delay transform and the foreground wedge}
\label{sec:delayfgndwedge}
This method relies on transforming the measured brightness temperature along each line of sight $T(p, \nu)$, into the conjugate delay domain yields $\tilde{T}(p, \tau)$ via a Fourier transform along the frequency axis, where the delay mode $\tau$ has dimensions of time. Because astrophysical foregrounds possess a smooth spectral dependence, their power is concentrated at low delay modes. Filtering or discarding these low-$\tau$ modes to isolate the cosmological 21\,cm signal represents a semi-blind foreground removal approach that avoids relying on explicit parametric models of the foreground spectra. The efficacy of map-based delay-space filtering has been demonstrated on simulated datasets, such as in \citet{2022MNRAS.517.4637P}.

The delay transform can be applied directly to raw visibilities to facilitate instrumental calibration \citep{2009AJ....138..219P} and foreground avoidance \citep{2012ApJ...752..137M, 2012ApJ...756..165P}. Visibilities are complex quantities characterized by a frequency-dependent amplitude and phase. Even for a source with a flat spectrum, geometric path delays introduce a baseline-dependent phase gradient across frequency.
Consequently, power from a spectrally flat sky, which concentrates at $\tau = 0$ in the line-of-sight delay transform $\tilde{T}(p, \tau)$, is dispersed across delay space for a visibility $\tilde{\mathcal{V}}_{ij}(\tau)$, extending up to the geometric horizon limit, depending on the baseline length:
\begin{equation}
    \tau_{\mathrm{max}} = \frac{|\vec{b}_{ij}|}{c}
\end{equation}
Furthermore, this delay-space power leakage is significantly more pronounced for interferometers composed of wide field-of-view antennas (such as dipoles) than for dish-based architectures \citep{2015ApJ...804...14T}.

\begin{figure}[htbp]
\hspace*{0.25\textwidth}
\includegraphics[width=0.40\textwidth]{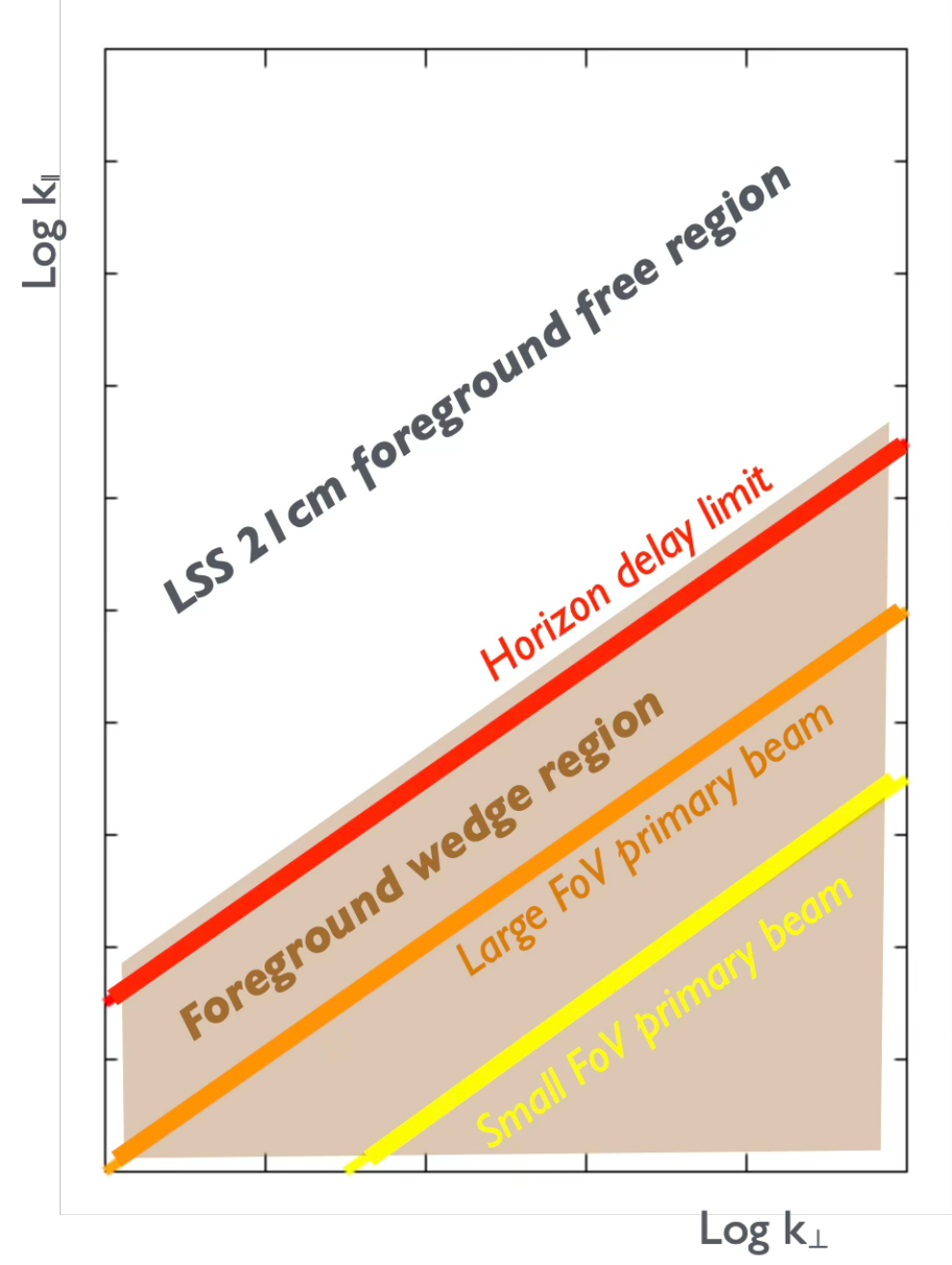}

\caption{Schematic representation of the foreground wedge region in the $(k_\perp, k_\parallel)$ plane, which is highly contaminated by foreground emission (illustration adapted from \citep{2012ApJ...752..137M}). 
 \label{fig:foregroundwedge}}
\end{figure}

In addition, chromatic primary beam effects and instrumental bandpass response induce power leakage from foregrounds beyond the intrinsic horizon delay limits. Nevertheless, one can estimate the 21\,cm power spectrum directly from visibilities in the delay domain. By discarding modes below the horizon delay threshold, $\tau \le \tau_\mathrm{max}(\vec{b}_{ij})$, it is possible to define a region in the spatial wavemode $(k_\perp, k_\parallel)$ plane that is virtually free of foreground contamination. Recognizing that delay $\tau$ maps to line-of-sight wavemode $k_\parallel$ and baseline length $|\vec{b}_{ij}|$ maps to transverse wavemode $k_\perp$, the foreground-dominated region forms a characteristic wedge-shaped domain defined by $k_\parallel \le k_\parallel^\mathrm{max}(k_\perp)$ \citep{2010ApJ...724..526D, 2012ApJ...752..137M, 2012ApJ...756..165P}. 
A schematic representation of the foreground wedge
is shown in Figure~\ref{fig:foregroundwedge}.  
As depicted, the wedge upper limit $k_\parallel^{limit} \sim \beta k_\perp$, has a slope $\beta$ increasing with the primary beam width, as 
shown in \citep{2015ApJ...804...14T}. 

\section{Observed 21\,cm signal by intensity mapping surveys} 
\label{sec:obs21cmsignal}

\subsection{Single-dish proof of concept surveys}
Conducted between 1997 and 2001 using the 64-meter Parkes Radio Telescope equipped with the 13-beam Parkes 21\,cm Multibeam Receiver \citep{1996PASA...13..243S}, the \HI Parkes All Sky Survey (HIPASS) covers a significant portion of both the Southern and Northern skies \citep{2004MNRAS.350.1195M}.
The survey spans a velocity range of $-1280 < cz < 12700 \, \mathrm{km/s}$, 
extending up to a maximum redshift $z \simeq 0.042$. A 2009 study \citep{2009MNRAS.394L...6P} successfully reported a positive cross-correlation between the 21\,cm signals from HIPASS data cubes and galaxies from the 6dF survey \citep{2009MNRAS.399..683J}.

Subsequently, Parkes was used in 2014 to observe two contiguous field of 2dF Galaxy Redshift Survey \citep{1999MNRAS.308..459F}, 
and accumulated 152 hours of data, covering $1300 \, \mathrm{deg^2}$. A clean 50\,MHz bandwidth centered at $ \sim 1320 \, \mathrm{MHz}$, corresponding to the redshift range  $0.057 < z <0.098$ was analyzed and cross-correlated with 2dF galaxies, obtaining a positive correlation with $5.7 \sigma$ significance, although a cross power spectrum amplitude much lower, by a factor 6, than the expected one  \citep{2018MNRAS.476.3382A}. A stronger correlation was measured for blue galaxies, compared to the red ones, confirming the gas depletion in massive galaxies. 

The fully steerable, 100-meter Green Bank Telescope (GBT)\footnote{Robert C. Byrd Green Bank Telescope \href{https://greenbankobservatory.org/about/telescopes/gbt/}{https://greenbankobservatory.org/about/telescopes/gbt/}}
has also been utilized to statistically detect the redshifted 21\,cm signal through cross-correlation with optical surveys. Observations were conducted between 2010 and 2015 using the 680-920 MHz prime focus receiver, which covers a redshift range of $ 0.6 < z < 1.0$.
An initial cross-correlation detection using DEEP2 \citep{2013ApJS..208....5N} fields was reported in 2010 \citep{2010Natur.466..463C}. 
This result was subsequently confirmed in 2013, using data from 190 hours of observations of $41 \,\mathrm{deg^2}$ of sky overlapping the WiggleZ survey \citep{2010MNRAS.401.1429D}, leading to a $\sim 7 \sigma$ detection of the cross-correlation on scales $0.05 \lesssim k \lesssim 0.8 \, h \mathrm{Mpc^{-1}}$ \citep{2013ApJ...763L..20M}. Translated into a measure of the \HI density at $z \sim 0.8$, the authors obtain $\Omega_{\mHI} b_{\mHI} r_{\mHI} \sim 0.43 \pm 0.07 \pm 0.04 \times 10^{-3}$. 
Furthermore, using the GBT radio observations alone, the 21\,cm brightness fluctuations at $z \sim 0.8$ was detected, as excess power in the autocorrelation power spectrum \citep{2013MNRAS.434L..46S}. 
More recently, a subset of the GBT data, comprising roughly 100 hours of observations over $\sim 100\, \mathrm{deg^2}$ was cross-correlated with the 
SDSS-IV eBOSS \citep{2016AJ....151...44D} and WiggleZ spectroscopic redshift catalogs to measure the \HI gas fraction at redshift $z \sim 0.8$ \citep{2022MNRAS.510.3495W}. Specifically, the authors constrained the product of the hydrogen mass-reduced density $\Omega_{\mHI}$, its bias $b_{\mHI}$, and the galaxy-neutral hydrogen correlation coefficient $r_{\mHI}$ to 
$\Omega_{\mHI} b_{\mHI} r_{\mHI} \sim 0.35-0.6 \times 10^{-3}$.

\subsection{CHIME}
\label{sec:chime}
CHIME (Canadian Hydrogen Intensity Mapping Experiment)\citep{2022ApJS..261...29C} is a custom-built, transit (drift-scanned) radio interferometer composed of four cylinder radio telescopes packed together, side-by side.  Each reflector is $20~\mathrm{m} \times 100~\mathrm{m}$ and is instrumented with a focal line of 256 dual-polarization feed antennas observing at 400–800 MHz ($2.55 \le z  \le  0.78$).  Besides pioneering 21\,cm IM, other science goals include searching for transient sources such as fast radio bursts and pulsars. It began taking first science data in late 2018. 

CHIME's first measurements of \HI came from cross-correlation (stacking) with objects from the eBOSS \citep{2023ApJ...947...16A} optical spectroscopic redshift survey.  The analysis is based on 102 nights of usable observations, acquired from January to November 2019, corresponding to a total of 521 hours of data.  RFI restricted the measurements to the frequency band  $587.5 - 800\, \mathrm{MHz}$, corresponding to the redshift range $0.8 < z < 1.4$.  The analysis focused on the North Galactic Cap (NGC) region, overlapping eBOSS. The eBOSS NGC footprint represents 2566 deg$^2$ for LRG (Luminous Red Galaxies), QSO (quasars), and ELG (Emission Line Galaxies).  There are a total of 39706 LRG's, corresponding to an effective redshift range $0.8<z<0.87$ and volume $V \sim 4 \times 10^8 \left( h^{-1} \mathrm{Mpc} \right)^3$,
and 94706 QSO's , $1<z<1.36$, corresponding to the volume  $V \sim 3 \times 10^9\, \left( h^{-1}  \mathrm{Mpc} \right)^3$ and 63381 ELG's over 550 deg$^2$, covering an effective redshift range $0.83<z<1.03$, 
and hence a cosmic volume $V \sim 2.6 \times 10^8\, \left( h^{-1}\mathrm{Mpc} \right)^3$.

After calibration and various data cleaning/flagging steps (RFI, sidereal day, baselines) and bright source subtraction, data are averaged by sidereal day. Signals from the feed antennas within each cylinder are combined with appropriate phasing (beamforming) to create 256 beams in the North-South direction.  These synthesized beams are then used to form visibilities for baselines in the East-West direction. Using a variant of m-mode map making, 
 sky maps are created for each frequency. This process neglects intra-cylinder baselines, which suffer from large cross-talk between the cylinder feed antennas, as well as large contributions from diffuse Galactic foregrounds. Foreground subtraction is performed by applying a delay transform to each sky direction, with a declination-dependent threshold.  Filtering the foregrounds in this way allows high significance detections of \HI at small, nonlinear scales, but reduces the sensitivity to the first few BAO peaks at larger scales. 
The cosmological scales being probed at $z \sim 1$ are $0.1 \lesssim k_\perp \lesssim 0.6 \, h\mathrm{Mpc}^{-1}$ and $0.25 \lesssim k_\parallel \lesssim 2 \,h\mathrm{Mpc}^{-1}$. 

The cross-correlation signal is obtained by stacking the maps around each source position, using the eBOSS source redshift, and is performed independently for each source category (LRG, QSO, ELG). These analyses allowed a measurement of  $A_{HI} = 10^3 \Omega_{HI} \left( b_{HI} + \langle f \mu^2 \rangle \right) $ ranging from 1.6 to 6 depending on the tracer and redshift, with the highest value for ELG. Using fixed nonlinear parameters such as the $b_\mHI$ value, the obtained $A_\mHI$ can be converted into an \HI density in the range $\Omega_\mHI = 0.8 - 1.2 \times 10^{-3}$, depending on the tracer, higher than other measurements in the same redshift range $0.8<z<1.4$. 

A similar approach (high-pass delay filtering of $k_{||} \lesssim 0.13\, \mathrm{Mpc}^{-1}$) was used for CHIME cross-correlation with eBOSS $\mathrm{Ly}\alpha$ forest spectra \citep{2024ApJ...963...23A}. Using 88 days of 400-500 MHz data ($1.8 \le z\le 2.5$), CHIME achieved a $9\sigma$ cross-correlation (stacking) detection of \HI -- the highest redshift detection of \HI emission so far.

The first auto-power \HI IM with CHIME \citep{2025arXiv251119620C,2026arXiv260325680C} demonstrated its ability to measure LSS without correlating with external templates. 
This analysis used 94 nights of observation, out of 102 nights of the same data set described above for the cross-correlation analysis, and was restricted to a $2200\, \mathrm{deg}^2$ sky region near the NGC and the frequency range from 608.2\,MHz to 707.8\,MHz ($1.01 \le z \le 1.34$), where RFI was lowest.  \HI auto-power was measured for $0.4 \,\mathrm{h} \mathrm{Mpc}^{-1} \le k \le 1.5 \,\mathrm{h} \mathrm{Mpc}^{-1}$  with a detection significance of $12.5\sigma$.   The \HI auto-power measurements are beginning to constrain the combination of the cosmic HI abundance ($\Omega_{HI}$) and the linear bias of \HI ($b_{HI}$):  $\Omega_{HI}^2(b_{HI}^2+\langle f \mu^2 \rangle$),   where $\langle f \mu^2 \rangle$ is a factor capturing the angular average of the Kaiser redshift-space distortion term over the wavenumbers  measured.
Several improvements were made over the pipeline used for cross-correlation detections, including  better identification and masking of RFI contamination  and improved beamforming in the NS direction to create achromatic beams and avoid frequency-dependent aliasing.  As with the earlier analyses, the auto-power analysis suppresses foreground by excluding intracylinder baselines and aggressively suppressing large line-of-sight modes (low $k_{||}$). 

These analysis choices
enable high significance detection of the 21\,cm signal at nonlinear
scales but essentially remove the BAO signal at larger scales. 
 Continued development of analysis techniques that control systematic effects will be required to measure larger scales.

\subsection{Tianlai}
\label{sec:tianlai}
Tianlai, meaning {\it Heavenly Sound} in Chinese, is an international 21\,cm Intensity Mapping project led by the National Astronomical Observatories of China (NAOC)\footnote{NAOC: \href{http://english.nao.cas.cn}{http://english.nao.cas.cn}}, 
with key contributions from the United States, France, and Canada. The collaboration operates two pathfinder instruments installed in a radio-quiet region of northwestern China: a cylindrical reflector array (TCA) and a dish array interferometer (TDA). Located near Hongliuxia in the Xinjiang province, approximately 500 km east of Urumqi, the observatory sits at the longitude 
$91^\circ 48' E$, latitude $44^\circ 09' N$, at an altitude of $ \sim 1500 \mathrm{m}$. 
While the feeds are designed to cover a broad frequency range of $400-1430 \mathrm{MHz}$, the current digitization and correlator electronics are restricted to a $100 \mathrm{MHz}$ bandwidth. This frequency band is tunable by adjusting the local oscillator and replacing the analog filters; to date, all observations have been conducted within the 700–800 MHz window, though there is a plan to switch to a low-$z$ survey in the 1320-1420 MHz band in the near future \cite{2022MNRAS.517.4637P}. 

The TCA consists of three cylinders, each measuring $40 \mathrm{m}$ in length and $15 \mathrm{m}$ in width, with instrumentation currently restricted to the central section of the focal line. Across the three cylinders, a total of 96 dual-polarization feeds collect 192 radio frequency (RF) signals. The digital correlator processes these sampled signals to compute $18 528$ visibilities, which are channelized into $896$ frequency channels at a frequency resolution of $122 \mathrm{kHz}$. The calibration method is presented in \cite{2019AJ....157...34Z}, and an end-to-end simulation is presented in \cite{2023RAA....23j5008Y}. The basic performance was analyzed in \cite{2020SCPMA..6329862L}. The array is also equipped with an transient backend capable of searching for fast radio burst (FRB) \cite{2024RAA....24h5010Y}. 
Some preliminary results including foreground mitigation and power spectrum analysis were presented in \citep{Deng2026}.   

The TDA consists of 16 on-axis dishes, each 6 meters in diameter, arranged in a concentric circular layout: a single central dish is surrounded by an inner hexagon of six dishes, which is further enclosed by an outer ring of nine dishes.
The dish layout was chosen among several options, optimizing spherical harmonic plane $(\ell, m)$ coverage \citep{2016MNRAS.461.1950Z}. 
Although the dishes are fully steerable, the array operates in transit mode, with all elements fixed and pointed toward a specific direction, usually along the meridian plane. Each dish features a dual-polarization receiver, generating a total of 32 RF signals that are sampled at 250 MSPS with 14-bit resolution. An FPGA-based correlator then computes 528 visibilities (496 cross-correlations and 32 autocorrelations) across 512 frequency channels at a 244\, kHz resolution. 
To date, several thousand hours of observations have been recorded, with a significant portion dedicated to the North Celestial Pole (NCP). A primary advantage of this dish array is its steerability toward specific declination bands; by tracking the NCP continuously, the array can achieve high sensitivity over a concentrated area of the sky. A detailed description of the Tianlai dish array, its performance, and preliminary results from deep NCP observations can be found in \citep{2021MNRAS.506.3455W,2026RAA....26i5013H}. A low-z survey \citep{2022MNRAS.517.4637P} is planned. 

An upgrade of the Tianlai experiment (called Tianlai II) is being made, with extensions to both the cylinder array and dish array, as well as the addition of  a new global 21\,cm spectrum measurement experiment. The new global spectrum experiment have two sites (G1, G2), which are located about 8 km East of the original site, with relatively flat ground. Conducting ground nets are installed at the G1 and G2 site, to facilitate testing different antennas and receivers of cosmic dawn 21cm global spectrum experiment.   

For the extension of the cylinder array, each of the existing three cylinders will get 16 additional dual polarization feeds, and the position of the existing feeds will also be adjusted, so that they will have longer baselines along the NS direction. Furthermore, three additional cylinders are built, each 20 meters long (NS) and 15 meters wide (EW). The center of cylinder D is located about (75 m W, 48 m S) of the center of the cylinder A (the original west cylinder), thus making longer baselines available for the intensity mapping survey. Additionally, two new cylinders (cylinder E, F) are placed some distance away from the original site, which are located at (8.857 km E, 3.235 km N), and (8.428 km E, 0.87km S)  respectively. Each of the new cylinders D, E, F are to be equipped with 16 dual polarization feeds, and serve as outtriggers for fast radio burst (FRB) locator. 

For the extension of the dish array, two additional dishes are added to the Northeast and Southeast of the original array, with baselines of about 100 m. The longer baselines will help improve the precision of the calibration of the dish array.

\subsection{MeerKAT}
\label{sec:meerkat}
MeerKAT (KAT stands for Karoo Array Telescope) is a radio array that serves as a pathfinder for the SKA - Mid array (Sec. \ref{sec:future}).  MeerKAT and SKA-Mid target 21-cm IM as one of many science goals, which also include galaxy evolution, transient phenomena, and tests of fundamental physics \citep{Jonas:2018Jr}.  MeerKAT has 64 off-axis dishes, each of diameter 13.5~m. 48 dishes are arranged in a core within a diameter of 1~km, with the full array within a diameter of 8 km.  MeerKAT's dishes will become part of the 197 - dish SKA-Mid, located in the Karoo Desert in South Africa.  These arrays are designed to operate both as radio interferometers and as arrays of independent single-dishes, simultaneously. The two observing modes provide unique opportunities for studying systematic effects.
The strategy for performing \HI IM measurements made in interferometer mode for small angular scales, single-dish mode for large-scales, and a combination of the two for intermediate scales is studied in detail in \citep{2015ApJ...803...21B}.  

Most MeerKAT 21\,cm IM data is obtained from two observing programs:  MIGHTEE and MeerKLASS.   MIGHTEE (MeerKAT International GHz Tiered Extragalactic Exploration) is a deep extragalactic survey, with a primary goal of studying galaxy evolution across cosmic time. It observes a relatively small area of sky very deeply, focusing on four well-studied extragalactic fields covering about 20 square degrees.  These observations are carried out in interferometric mode, tracking a target sky direction.  Multiple array pointings are then combined by mosaicking.

The first detection of the \HI auto-power spectrum with MeerKAT \citep{paul_direct_2026} was made with 96 hrs of data obtained in interferometric mode on a single field during pilot observations for the MIGHTEE program.   Data were analyzed at frequencies centered at 986 MHz ($z \sim 0.44$) and 1077.5 MHz
($z \sim 0.32$).   Foregrounds were managed by foreground avoidance, removing the foreground wedge in the 2-D $(k_\perp, k_\parallel)$ power spectrum plane, using the delay transform (see section \ref{sec:delayfgndwedge}).  The 3D average k-range is $ k \sim  1 - 5$, higher in $k_\perp (\lesssim 20)$, lower in $k_\parallel (\lesssim 2-3)$. The statistical significance is about $3.5 \sigma$, and up to $6-9 \sigma$, depending on the analysis method.

MIGHTEE IM observations were also used to detect \HI by stacking interferometer maps on galaxies from the COSMOS field at $0.23 < z < 0.49$ \citep{sinigaglia_mightee-hi_2022}.  16 hours of observation were obtained on a 1.5 $\mathrm{deg^2}$ field.  Stacking on 9023 star-forming galaxies with spectroscopic redshifts, selected from the COSMOS catalog, resulted in a detection with statistical significance of about $5 \sigma$.  In terms of observing approach and k-space sensitivity, this analysis is similar to that of GMRT (Sec. \ref{sec:gmrt}.).

Other MIGHTEE measurements \citep{2025MNRAS.541..476M} are upper limits for \HI auto- and cross-power (with the COSMOS field), at $0.5 < k < 10 \, \mathrm{Mpc^{-1}}$.   Recently, \citep{2026arXiv260223055T}, claims  a statistically significant  detection of the \HI auto-power spectrum at $z \sim 0.05$ on scales $ 3 \leq k \leq 20 
\, \mathrm{Mpc}^{-1}$ with 17.5 hours of observations on a single field. 

MeerKLASS (MeerKAT Large Area Synoptic Survey) \citep{2016mks..confE..32S} has a primary objective of mapping the large-scale structure of the universe using 21\,cm IM. Compared to MIGHTEE, access to lower k modes is obtained by operating in single-dish mode (multi-dish autocorrelation). The dishes scan back-and-forth in azimuth at constant elevation to minimize ground pickup and atmospheric fluctuations, with noise
diodes firing at regular intervals to provide a stable reference
for gain calibration. This approach ensures reasonably uniform
sky coverage while maintaining instrumental stability.
In parallel, the interferometric visibilities from the constant elevation scans are also recorded and processed 
through the on-the-fly (OTF) mapmaking pipeline, which produces high angular resolution ($\sim 25 \mathrm{arcsec}$ in the UHF band, 544-1088 MHz)  
to support commensal science \citep{2025arXiv251211978C}.

 As reported in \citep{2023MNRAS.518.6262C}, MeerKLASS detected correlated clustering between MeerKAT intensity maps and galaxies from the WiggleZ Dark Energy Survey \citep{2010MNRAS.401.1429D}.  A  $\sim 200\,\mathrm{deg}^2$ region overlapping the WiggleZ survey was observed during 10.5 hours in the single-dish mode. A fraction of the available L-band data in the frequency range $\left[973, 1014 \right] \,\mathrm{MHz}$, corresponding to $0.4<z<0.459$, was used to create 3D radio sky maps. These were then convolved to a common lower angular resolution beam to minimize mode mixing due to the frequency-dependent beam, and then subjected to a 
PCA foreground cleaning phase (see section \ref{sec:fgndclean}), by removing the first 30 modes. The cross-power spectrum was detected at $7.7\sigma$ for $0.06 \lesssim k \lesssim 0.3$, complementing the higher k range of MeerKAT interferometry. This detection provides a measurement of $ \Omega_{HI} b_{HI} r \simeq 0.86 \times 10^{-3} $  around redshift $z \sim 0.43$ at $k_{eff} \simeq 0.13 \,h \mathrm{Mpc}^{-1}$.  

An additional $\sim 4 \sigma$ detection of the cross-power spectrum for $0.39 < z < 0.46$ has been reported in \citep{2025MNRAS.537.3632M}, by combining the MeerKLASS intensity maps with the GAMA \citep{2015MNRAS.452.2087L} redshift survey data. This analysis of 62 hours of data obtained from 2021 single-dish mode observation of 236 deg$^2$, compared to 10.5 hours of the pilot survey discussed above, shows improved data quality, along with an improved analysis pipeline, mostly in the calibration procedure.  The foreground cleaning approach is similar to the one mentioned above (PCA after common beam convolution). 
The \HI signal has also been detected in the same data set through stacking \cite{2025ApJS..279...19C} on 2269 galaxies from the GAMA survey. 
 An extensive discussion of foreground removal in the context of MeerKAT single-dish observations is presented in \citep{2025A&A...703A.222C} and applied to a subset of the MeerKLASS data. 

Using this same 62 hours of data, in the same frequency band ($ 971.2 < \nu <1023.6 \mathrm{MHz}$) MeerKLASS has obtained a measurement of the \HI auto-correlation signal 
in the wavenumber range $0.095 < k < 0.245 \, h \mathrm{Mpc}^{-1}$ and for $0.39 < z < 0.46$ \citep{2026arXiv261012158M}.


\subsection{FAST}
\label{sec:fast}
The Five-hundred-meter Aperture Spherical radio Telescope (FAST) \citep{2011IJMPD..20..989N} is currently the world's largest single-dish radio telescope. It features a fixed $500\,\mathrm{m}$ spherical primary reflector; however, its active illuminated aperture forms a $300\,\mathrm{m}$ paraboloid at any given time, enabling observations up to a zenith angle of $40^\circ$. With an aperture efficiency of $\approx 70\%$, its effective collecting area reaches $A_\mathrm{eff} \approx 5 \times 10^4\,\mathrm{m^2}$, yielding an angular resolution of $\sim 2.9\,\mathrm{arcmin}$ at $1.4\,\mathrm{GHz}$. 

Equipped with a 19-beam L-band receiver \citep{2020RAA....20...64J} covering the frequency range $1.05\text{--}1.45\,\mathrm{GHz}$, FAST is a highly competitive facility for $21\,\mathrm{cm}$ intensity mapping (IM). This frequency band enables IM surveys of \HI up to redshift $z \approx 0.35$. As demonstrated by \citet{2020MNRAS.493.5854H}, a FAST $21\,\mathrm{cm}$ IM survey significantly outperforms an \HI galaxy survey for an equivalent integration time; specifically, IM achieves sufficient sensitivity to detect the $21\,\mathrm{cm}$ power spectrum $P_{21}(k)$ across cosmological scales $0.01 \lesssim k \lesssim 1\,h\,\mathrm{Mpc}^{-1}$ relevant for baryon acoustic oscillations (BAO) up to $z \approx 0.35$, whereas a galaxy survey with the same instrument is limited to $z \lesssim 0.15$.
But currently, the usable frequency range for 21\,cm intensity mapping is limited to the 1320-1420 MHz and 1050-1100 MHz bands, as the frequencies beyond these two bands are heavily contaminated by radio frequency interference (RFI).

Several surveys in drift-scan mode are being carried out with FAST. The telescope is pointed toward a given direction $(\alpha_0, \delta_0)$, and a stripe of sky at the fixed declination $\delta_0$ is observed as the earth rotates. FAST All Sky \HI  survey (FASHI) \citep{2024SCPMA..6719511Z} plan to observe with the L-band receiver all the sky in the declination range $-14^\circ < \delta < 66^\circ$ accessible to FAST, representing around $\sim 22000 \mathrm{deg^2}$. 
In a first data release of this blind \HI galaxy survey, the FASHI team has identified 41741 sources over $\sim 7600 \mathrm{deg^2}$ \citep{2024SCPMA..6719511Z}, to be compared with 31502 objects detected by the Arecibo ALFALFA survey \citep{2018ApJ...861...49H} over $\sim 7000 \mathrm{deg^2}$.  Commensal Radio Astronomy FasT (CRAFTS) \citep{2018IMMag..19..112L} is another wide area survey, being carried out with the 19 beam L-band receiver in drift-scan mode, which combines \HI imaging and pulsar search. 

The FAst neuTral HydrOgen intensity Mapping ExpeRiment (FATHOMER) \citep{2023ApJ...954..139L} is a drift-scan survey dedicated to neutral hydrogen \HI IM. Pilot observations were conducted over several nights in 2021 through 4-hour drift scans near a target declination of $\delta_0 = 26^\circ$, centered at $\mathrm{RA} = 11^\mathrm{h}00^\mathrm{m}$ and covering an area of $\sim 60\,\mathrm{deg^2}$. The survey team successfully calibrated and RFI-cleaned these data, generating maps across two usable frequency sub-bands: $1050\text{--}1150\,\mathrm{MHz}$ and $1323\text{--}1450\,\mathrm{MHz}$. 

The reconstructed maps, discretized into voxels of $2.95\,\mathrm{arcmin^2} \times 28\,\mathrm{kHz}$, exhibit root-mean-square (RMS) noise fluctuations of $36.0\,\mathrm{mK}$ and $25.9\,\mathrm{mK}$ in the low- and high-frequency bands, respectively. These values are slightly higher than expected noise levels calculated assuming a system temperature of $\mTsys = 20\,\mathrm{K}$. Utilizing this same map dataset, \citet{2026ApJ..1001..204S} identified 702 \text{\HI} sources, compared to the 331 previously detected by the ALFALFA survey within the overlapping footprint. The FATHOMER catalog demonstrates a better sensitivity and extends to greater redshifts, beyond $cz > 20000\,\mathrm{km\,s^{-1}}$.

There is a plan to construct an array of smaller radio antennas (40\, m aperture) within a radius of 5\, km around the FAST \cite{2024AstTI...1...84J}, to take advantage the unprecedented sensitivity of FAST while achieving a high angular resolution with the interferometer array. The observations of such an array can be carried out with both cross-correlation and autocorrelation visibilities. The autocorrelation visibilities of these dish antennas can be used for \HI intensity mapping as well, and the interferometry measurements would be useful in calibrating the gains of the receivers. 

\subsection{GMRT}
\label{sec:gmrt}
The Giant Metrewave Radio Telescope (GMRT)\footnote{GMRT : \href{https://www.gmrt.ncra.tifr.res.in/index.html}{https://www.gmrt.ncra.tifr.res.in/index.html}} is a high-sensitivity interferometer with 30 antennas, each with a 45\,m diameter reflector. With baseline lengths up to 25\,km, GMRT can reach resolutions of a few arcseconds for observation frequencies near 1 GHz. As such, GMRT is not suited to 21\,cm IM observations as discussed here. However, \citet{GMRT_stacking_zone} have used the Upgraded GMRT (uGMRT) to observe a small sky area of 1.2\, $\mathrm{deg^2}$ overlapping with the DEEP2 \citep{2013ApJS..208....5N} galaxy survey, in the frequency band 580-820\,MHz, corresponding to the redshift range $0.74 \lesssim z \lesssim 1.45$. Using a position and frequency stacking technique over 7653 DEEP2 galaxies, they  obtain a stacked 21\,cm \HI emission signal with more than $4 \sigma$ significance, corresponding to an average \HI density $\Omega_\mHI \sim 0.5 \times 10^{-3}$  at $z \sim 1$ \citep{GMRT_stacking_zone}. A refined version of this analysis, using 510 hours of GMRT observations over $\sim 2 \, \mathrm{deg^2}$ is presented in \citep{2022ApJ...937..103C}, with \HI emission detected at $7.1 \sigma$, stacked over 11419 blue star-forming galaxies.   

\begin{figure}[H]
\hspace*{0.25\textwidth} 
\includegraphics[width=0.5\textwidth]{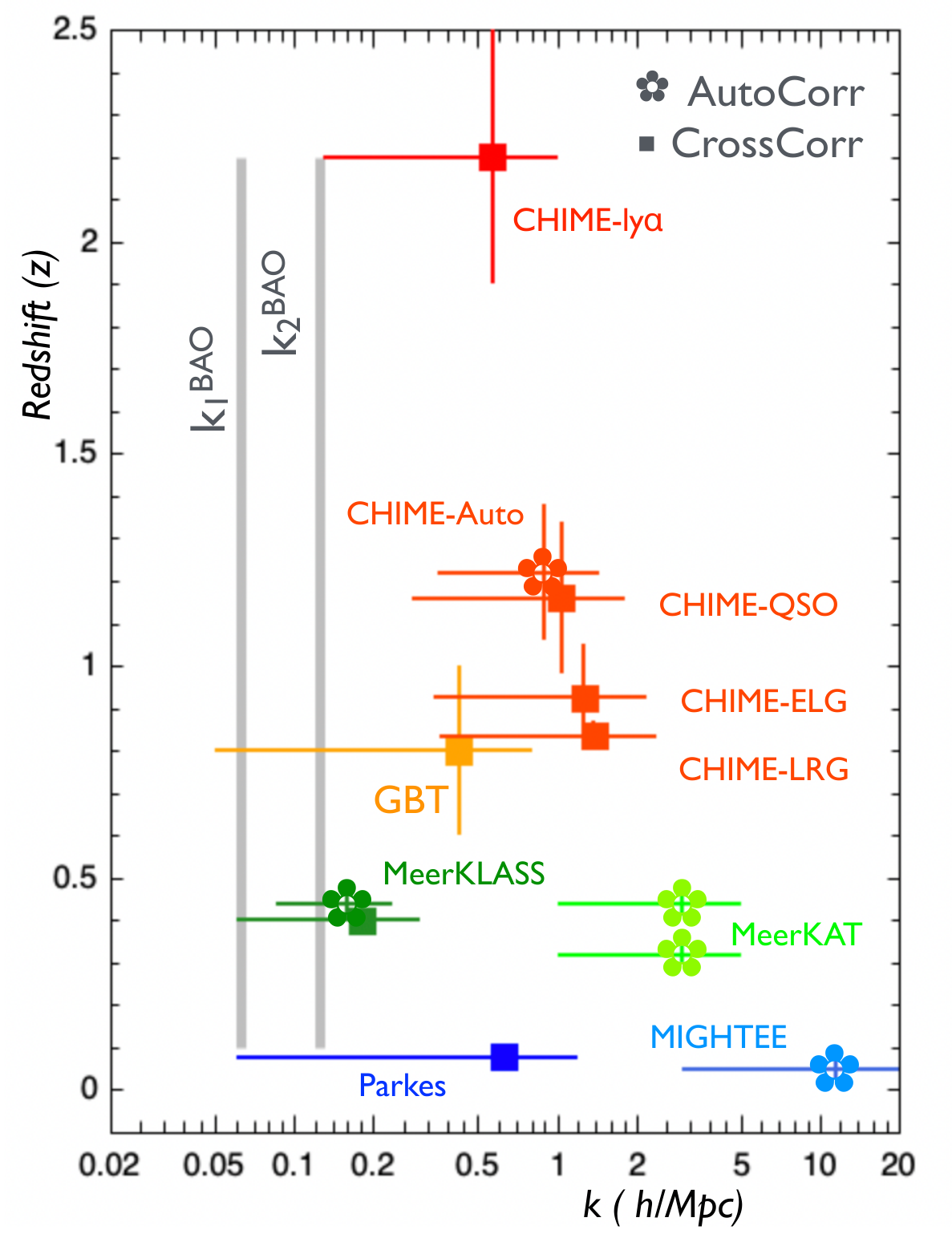}
\caption{Cross-correlation, and \HI autocorrelation measurements by Parkes, GBT, MeerKAT and CHIME, as listed in Table \ref{tab:cxsummary}, in the wavenumber, $k$ and redshift, $z$ plane. 
The approximate position of the  first two BAO wiggles in the power spectrum ($k_1^\mathrm{BAO}, k_2^\mathrm{BAO}$) are shown as grey vertical lines. 
The shown positions corresponds to $(2 \pi / r_s, 4 \pi / r_s)$ with the sound horizon at recombination $r_s \simeq 145\, \mathrm{Mpc} \simeq 100 \, h^{-1} \mathrm{Mpc}$ as determined by Planck \cite{2021A&A...652C...4P}, while the real BAO peak positions would be shifted by 10-20\% toward higer k-value. Note also that the position of the measurements corresponding to CHIME-ELG and CHIME autocorrelation pair (marked as as (d)-ELG, (h) in  Table \ref{tab:cxsummary}) and MeerKLASS auto and cross correlation pair (marked as as (c), (i) in  Table \ref{tab:cxsummary}) have been shifted slightly for clarity. \label{fig:cxsummary}}
\end{figure}

A summary of some of the results from the 21\,cm surveys described here appears in Table \ref{tab:cxsummary} and the associated Fig. \ref{fig:cxsummary}. 

\begin{table}
\begin{tabular}{|l|c|c|c|c|c|}
\hline
    & redshift & $\Omega_\mathrm{surv}$ & $t_\mathrm{surv}$ & $\delta \theta$ & k-range  \\
    \hline
    Parkes $\otimes$ 2dF (a) & $0.057 - 0.098 $ & 1300 & 152 & 12' & $ 0.06 - 1.2 $ \\
    GBT $\otimes$ WiggleZ (b) & $0.6 - 1 $ & 41 & 190 & 13' &  $0.05 - 0.8$   \\
    MeerKLASS $\otimes$ WiggleZ (c) & $0.4 - 0.459 $ & 200 & 10.5 & 76' &  $ 0.06 - 0.3 $ \\
    CHIME $\otimes$ eBOSS-LRG (d) & $0.8 - 0.87$ & 2566 & 521 & 14.5' & $0.36 - 2.4$  \\
    CHIME $\otimes$ eBOSS-ELG (d) & $0.8 - 1.05 $ & 550 & 521 & 18' &  $0.33 - 2.2$  \\
    CHIME $\otimes$ eBOSS-QSO (d) & $1 - 1.36$ & 2566 & 521 & 21' & $0.27 - 1.8$ \\
    CHIME $\otimes$ eBOSS-Ly$\alpha$ (e) & $1.8 - 2.5$ & $\sim$4000 & $\sim 1000$ & 40' & $0.13 - 1$ \\
    \hline
    MIGHTEE (f) & $0.02 - 0.07$ & $\sim 3$ & 17.5 & 0.7' & $3 - 20$ \\
    MeerKAT-Interf (g) & $0.29 - 0.35$ & $\sim 6$ & 96 & 1' & $\sim 1-5 $ \\
    MeerKAT-Interf (g) & $0.41 - 0.47$ & $\sim 6$ & 96 & 1.1' & $\sim 1 - 5$ \\
    CHIME-Auto (h) & $1.01 - 1.34$ & 2200 & $\sim 500$ & 21' & $0.4 - 1.5$ \\
    MeerKLASS (i) & $0.39 - 0.46$ & 236 & $62$ & 76' & $0.095-0.245$ \\ 
    \hline 
\end{tabular}
    \caption{Characteristics of some of the measured \HI - optical  cross-correlations (top rows, (a-e)) and 21\,cm autocorrelations (bottom rows, (f-i)) discussed here. The second column gives the analysed redshift range, while $\Omega_\mathrm{surv}$ is the surveyed sky area in $\mathrm{deg^2}$ and $t_\mathrm{surv}$ is the total survey duration in hours. $\delta \theta$ is an estimate of the angular resolution in arcmin, computed from the reflector diameter or the array size as $(1+z) \lambda_{21}/D$. The last column gives the probed wavenumber $k$ range, in $h \mathrm{Mpc}^{-1}$, as stated in the corresponding references, or estimated by the authors. (a)  \citep{2018MNRAS.476.3382A}; (b) \citep{2013ApJ...763L..20M}; (c)  \citep{2023MNRAS.518.6262C}; (d) \citep{2023ApJ...947...16A}; (e) \citep{2024ApJ...963...23A}; (f) \citep{2026arXiv260223055T}; (g) \citep{paul_direct_2026}; (h)  \citep{2026arXiv260325680C}; (i) \citep{2026arXiv261012158M}. \label{tab:cxsummary}} 
\end{table}

\section{Future surveys and instruments} 
\label{sec:future}

 HIRAX (Hydrogen Intensity and Real-time Analysis eXperiment) \citep{crichton_HIRAX2022} is a compact interferometric array composed of $6\text{ m}$ dish antennas.  Sharing significant technological heritage with CHIME, it will observe the same 400 to 800\,MHz band, corresponding to the redshift range of $2.55 > z > 0.775 $.  The initial 256-dish configuration provides a total collecting area of approximately $7200\text{ m}^2$. To minimize mutual coupling between antennas, HIRAX uses short focal length, on-axis dishes (f/D = 0.23), where the dual-polarization feed antennas are located just below the rim of each dish. HIRAX is currently under construction at the South African Radio Astronomy Observatory (SARAO) in the Karoo Desert, a few kilometers from the MeerKAT site, and is designed to survey over $\sim 15,000$ square degrees of the Southern Sky.  Future expansion is planned to the final, 1024 element array.

CHORD (Canadian Hydrogen Observatory and Radio Transient Detector) \citep{2026arXiv260709374T} builds on lessons learned from CHIME and inherits much of CHIME's core technology, particularly its custom electronics and FX correlator. However, departing from CHIME's cylindrical reflector design, CHORD's core array comprises 512 dish antennas ($D = 6\text{ m}$), providing a total collecting area of $14{,}400\text{ m}^2$. Similar to HIRAX, CHORD uses short focal length, on-axis dishes (f/D $\sim 0.21$), placing the focal plane slightly below
the plane of the dish rim.  Its ultra-wideband receivers cover $300\text{--}1500\text{ MHz}$ ($3.7 > z > - 0.05 $), extending beyond CHIME's redshift range. Additionally, wide-field outrigger stations positioned at long baselines from the main CHIME/CHORD site will provide high-precision localization for Fast Radio Bursts (FRBs).

BINGO (Baryon acoustic oscillations from Integrated Neutral Gas Observations) is a dedicated $21\text{cm}$ intensity mapping instrument currently under construction in Brazil, utilizing an innovative transit-mode, single-dish optical design \citep{2013MNRAS.434.1239B}. The system features an off-axis primary reflector with a surface area of approximately $1600\text{ m}^2$ ($D \sim 50\text{ m}$) that illuminates a focal plane populated with $28$ feed horns via a secondary mirror. The entire optical setup is fixed in place, observing a $15^\circ$-wide declination strip centered at $\delta = -15^\circ$. Operating across the $980\text{--}1260\text{ MHz}$ frequency range, BINGO will probe neutral hydrogen across the redshift interval $0.13 < z < 0.45$ \citep{2022A&A...664A..14A}.

QTT (QiTai radio telescope, 110 m aperture), HRT (the Huadian Radio Telescope, 120 m aperture), and JRT (the Jingdong Radio Telescope, 120 m aperture) are three large aperture radio telescopes with steerable dishes, under construction or being planned in China. Once operational, these telescopes are also quite suitable for conducting \HI intensity mapping surveys \cite{2025arXiv251118478P}. 

The Square Kilometre Array Observatory (SKAO) consists of two complementary instruments operating across different frequency bands: SKA-Low, located in Australia, and SKA-Mid, based in South Africa. 

In its currently planned AA* phase, SKA-Mid comprises 144 reflectors, combining 80 new 15-meter dish antennas with 64 existing 13.5-meter MeerKAT precursor antennas, while a total of 197 antennas are foreseen for the AA4 stage. 
Half of these dishes form a dense core spanning $\sim 1\text{ km}^2$, while the remainder are arranged along three spiral arms with baselines reaching up to $108 \text{ km}$ in AA* and $159\text{ km}$ in AA4 \footnote{SKAO Staged Delivery, Array Assemblies and Layouts document, June 2025, \href{https://zenodo.org/records/16951020}{https://zenodo.org/records/16951020} }.
The Science verification is scheduled to start in 2031 for the SKA-MID in its AA* configuration, to be followed by a shared risk operational cycle, in 2031-2032
\footnote{SKAO Timeline to Science (March 2026) : \href{https://www.skao.int/en/science-users/647/timeline-science}{https://www.skao.int/en/science-users/647/timeline-science}}.
Operating between $350\text{ MHz}$ and $15\text{ GHz}$, each antenna focal plane houses multiple receivers, though only one is active at a time. For post-Epoch of Reionization (EoR) intensity mapping, two bands are particularly relevant: the L-band receiver (Mid-2, $950\text{--}1700\text{ MHz}$, $0 < z < 0.35$) and the Mid-1 receiver, which extends the MeerKAT UHF band ($350\text{--}1050\text{ MHz}$, $0.35 < z < 3$). Compared to the Jansky VLA in the United States, SKA-Mid offers a $4\times$ improvement in angular resolution, a $5\times$ boost in sensitivity, and a $50\times$ increase in survey speed.

While SKA-Low ($50\text{--}350\text{ MHz}$) targets mainly the Cosmic Dawn and the EoR ($6 < z < 30$), SKA-Mid will map the LSS across different cosmic epochs. Individual galaxy detections in the $21\text{cm}$ line will yield a vast catalogue of \HI galaxies with precise redshifts up to $z < 0.4$ 
\citep{2020PASA...37....7S}. However, detecting individual \HI clouds at higher redshifts remains challenging, yielding only a sparse population of extremely bright galaxies up to $z \sim 1$. To bridge this gap, SKA-Mid Intensity Mapping surveys will probe the LSS and measure the Baryon Acoustic Oscillation (BAO) scale up to $z = 2$ or $z = 3$ \citep{2017MNRAS.466.2736V}. 

Beyond second-generation dedicated IM instruments currently under construction (e.g., CHORD and HIRAX), more ambitious concepts like PUMA have been proposed as Stage~II or Stage~III 21\,cm IM experiments. The PUMA concept \citep{2019BAAS...51g..53S} envisions a close-packed interferometric array comprising up to $32,000$ dishes with receivers spanning $200\text{--}1100\text{ MHz}$. This design will enable surveys over a vast cosmological volume across the redshift range $0.3 < z < 6$, holding the promise of a high scientific yield \citep{2018arXiv181009572C}.  

\section{Summary} 
\label{sec:summary}
We have witnessed tremendous progress in post-EoR 21\,cm IM over the last decade, marked by the commissioning and operation of several large radio telescopes, as well as large IM surveys conducted with general-purpose instruments. Several groups, using different instruments, have detected the cross-correlation of the 21\,cm extragalactic radio signal with optical redshift surveys at high statistical significance, as discussed in Section~\ref{sec:obs21cmsignal}. These remarkable results demonstrate that the observation strategies and data analysis methods developed to date can indeed cope with overwhelming Galactic synchrotron and point-source foregrounds. More recent claims of the \HI autocorrelation signal detection by several groups represent a major breakthrough and milestone in the field (Table~\ref{tab:cxsummary}).

Nevertheless, outstanding questions remain to be answered before declaring that the method has lived up to its promises. On one hand, the published results often rely on a small fraction of the total accumulated data, reflecting the immense challenges faced when developing data processing and analysis pipelines. The ability of packed interferometer arrays observing in transit mode to reach the extremely stringent calibration requirements for IM surveys has not yet been fully demonstrated and remains an open question. Lessons learned from first-generation transit array instruments tend to indicate that dish arrays with optical designs optimized to minimize sidelobes and cross-coupling may be a more optimal technical solution than cylindrical reflectors, despite having a significantly narrower primary beam. Finally, the data processing and foreground subtraction methods used so far are inefficient at large scales, and the prospect of detecting the LSS signal at BAO scales remains elusive.

The second-generation arrays, such as HIRAX and CHORD, presented in Section~\ref{sec:future}, will likely address these challenges and hopefully pave the way for PUMA-like third-generation instruments.


\authorcontributions{The authors shared equally in the effort and responsibility of gathering the information from previously published work and summarizing it for this paper. All authors have read and agreed to the published version of the manuscript. 
}

\funding{
This work was produced by Fermi Forward Discovery Group, LLC under Contract No. 89243024CSC000002 with the U.S. Department of Energy, Office of Science, Office of High Energy Physics. Publisher acknowledges the U.S. Government license to provide public access under the DOE Public Access Plan
(\url{https://www.energy.gov/doe-public-access-plan}. 
XLC is supported by the National Natural Science Foundation of China (NSFC) under grants Nos.12361141814, 12421003, by China's Space Origins Exploration Program Nos. GJ11010405, and by the Specialized Research Fund of State Key Laboratory of Radio Astronomy and Technology.
}

\dataavailability{We did not produce or directly use any data or software for this work. Interested readers should refer to the listed references for the availability of the corresponding data or software. 
}

\acknowledgments{We acknowledge discussions with Ue-Li Pen, Olivier Perdereau, Jeff Peterson, and our colleagues in the Tianlai Collaboration.}

\conflictsofinterest{
The authors declare no conflicts of interest.
} 



\abbreviations{Abbreviations}{
The following abbreviations are used in this manuscript:
\\

\noindent 
\begin{tabular}{@{}ll}
BAO & Baryon Acoustic Oscillations\\
CDM & Cold Dark Matter model\\
$\Lambda$CDM & Lambda Cold Dark Matter model\\
CMB & Cosmic Microwave Background\\
EoR & Epoch of Reionization\\
FoV & Field of View\\
\HI & Neutral hydrogen gas\\
IM & Intensity Mapping\\
LIM & Line Intensity Mapping\\
LSS & Large-Scale Structure\\
RF & Radio Frequency\\
RSD & Redshift-Space Distortions\\
VLBI& Very Long Baseline Interferometry\\
WIM & Warm Ionized Medium\\

\end{tabular}
}




\begin{adjustwidth}{-\extralength}{0cm}

\reftitle{References}


\bibliography{refsim21.bib}


%


\PublishersNote{}
\end{adjustwidth}
\end{document}